\documentclass[namedreferences]{solarphysics}
\usepackage[optionalrh,solaenum]{spr-sola-addons} 
\usepackage{graphicx}                    
\usepackage{courier}                    
\usepackage{amssymb}                    
\usepackage{color}                       
\usepackage{url}                         

\begin{document}
\begin{article}

\begin{opening}

\title{Influence of the Solar Global Magnetic Field Structure Evolution on CMEs}

%
\author{Irina~\surname{A. Bilenko}  
       }

%
 \runningauthor{Bilenko I.A.}
 \runningtitle{Solar Global Magnetic Field Structure Evolution and CMEs}

%
  \institute{$^{1}$ Moscow M.V. Lomonosov State University, Sternberg Astronomical
  Institute, Universitetsky pr.13, Moscow 119992, Russia
                     email: \url{bilenko@sai.msu.ru}      
           }

\begin{abstract}

The paper considers the influence of the solar global magnetic
field structure (GMFS) cycle evolution on the occurrence rate and
parameters of coronal mass ejections (CMEs) in cycles 23-24. It
has been shown that over solar cycles, CMEs are not distributed
randomly, but they are regulated by evolutionary changes in the
GMFS. It is proposed, that the generation of magnetic Rossby waves
in the solar tachocline results in the GMFS cycle changes. Each
Rossby wave period favors a particular GMFS. It is proposed that
the changes in wave periods result in the GMFS reorganization and
consequently in CME location, occurrence rate, and parameter
changes. The CME rate and parameters depend on the sharpness of
the GMFS changes, the strength of the global magnetic field and
the phase of a cycle.

\end{abstract}

%
\keywords{Solar Cycle; Magnetic fields; Oscillations; Coronal mass
ejections.}

\end{opening}

%
 \section{Introduction}  

CMEs are the most energetic solar activity phenomena. They have
speeds up to $\sim10^3$ km s$^{-1}$, total mass of $\sim10^{16}$
g, and energy reaching $\sim 10^{23} - 10^{25} $ erg
(\opencite{Howard1985}; \opencite{Cyr2000};
\opencite{Gopalswamy2006}). CMEs are considered to be associated
with large-scale, closed magnetic field structures in the solar
corona (\opencite{Hundhausen1993}; \opencite{Munro1979};
\opencite{Chen2000}; \opencite{Forbes2006};
\opencite{Gopalswamy2006}). They are caused by loss of equilibrium
of the pre-existing magnetic structure \cite{Schmieder2006}.
Significant parts of the solar atmosphere are involved in a CME.
CMEs are known to be the main drivers of space weather
\cite{Schwenn2006}. Ejected coronal plasma may cause strong
geomagnetic storms. CME source regions can be identified with
different types of large-scale
structures(\opencite{Hundhausen1993}; \opencite{Khan2000};
\opencite{Bemporad2005}; \opencite{Zhou2006}). CMEs may be caused
by the instability or lack of equilibrium in coronal loops
(\opencite{McAllister1996}; \opencite{Zhou2006};
\opencite{Lara2008}; \opencite{Harrison2010}). Some observations
have shown that CMEs are associated with streamers
(\opencite{Illing1986}; \opencite{Steinolfson1988};
\opencite{Hiei1993}; \opencite{Hundhausen1993};
\opencite{Subramanian1999}; \opencite{Floyd2014}). Other
investigators relate CMEs to coronal holes (\opencite{Hewish1986};
\opencite{Bilenko2009}) or sigmoid magnetic field structures
\cite{Sterling2000}. The initiations of CMEs are often found to be
related to the other solar activity, e.g., active regions (ARs),
flares, filaments/prominences, streamers, and coronal holes or in
different combinations.

On the whole, the CME activity tends to track a solar cycle
(\opencite{Webb1991}; \opencite{Webb1994}; \opencite{Hildner1976};
\opencite{Cyr2000}; \opencite{Gopalswamy2006};
\opencite{Cremades2007}; \opencite{Robbrecht2009a};
\opencite{Gerontidou2010}; \opencite{Lamy2014}), but it differs in
a significant way from that of the small scale solar activity
phenomena due to the fact it is more in line with the evolution of
the global magnetic field (\opencite{Li2009};
\opencite{Bilenko2012}). The changes in the distribution of CME
latitudes do not correspond to those related to small scale
magnetic structures such as sun spots or $H\alpha$ flares; they
resemble those related to large-scale magnetic structures, such as
prominences and bright coronal regions \cite{Hundhausen1993}.
During cycle 23 the CME activity shows a significant peak delay
with respect to the AR cycle \cite{Robbrecht2009a}. Large-scale
solar magnetic fields have a significant effect on the
characteristics and propagation of CMEs \cite{Fainshtein2010}.
Different CME parameters have different behavior in solar cycle
maxima and minima (\opencite{Hundhausen1993};
\opencite{Hundhausen1994}; \opencite{Vourlidas2010};
\opencite{Vourlidas2011}; \opencite{Bilenko2012}). This reflects
their association with different shapes and orientation of the
closed structures of the solar magnetic field at different phases
of a solar cycle \cite{Bravo1998} and the influence of the solar
global magnetic field cycle evolution (\opencite{Bilenko2012};
\opencite{Petrie2013}). The changes in the domination of the
sectorial and zonal structures of the solar global magnetic field
influence the CME rate and parameters. During the zonal structure
domination, the solar minima phases, CMEs have lower occurrence
rate and parameters on average. When sectorial structures begin to
dominate at the rising phase of solar activity, the sharp increase
in CME daily rate and parameters is observed. The latitudinal
distribution and the statistics of CME parameters are also
different for periods of zonal and sectorial structure domination
\cite{Bilenko2012}. In \inlinecite{Petrie2013} it was shown that
the rate of solar eruptions was higher for years 2003-2012 than
for years 1997-2002. This was explained by the weakness of the
late-cycle 23 polar fields and the influence of the changes in the
polar fields on the global coronal field structure.

Comparing the occurrence rates of CMEs with the long-term
evolution of the global white light coronal density distribution
\inlinecite{Sime1989} concluded that CMEs arise from pre-existing
magnetic structures which become stressed by the global magnetic
field rearrangement to the point of instability. CMEs are believed
to be a consequence of the coronal field rearrangement due to a
loss of stability of the magnetic field \cite{Forbes2000}. CMEs
are the result of a global magnetohydrodynamic process and
represent a significant restructuring or reconfiguring of the
global coronal magnetic field (\opencite{Harrison1990};
\opencite{Low1996}). The topology changes from closed to open
magnetic field configuration result in CMEs \cite{Wen2006}. And
vice versa CMEs can have influence on the coronal magnetic field
reconfiguration \cite{Liu2009}. \citeauthor{Low1996}
(\citeyear{Low1996}, \citeyear{Low2001}) suggested that CMEs can
be a basic mechanism of coronal magnetic field reconfiguration.
However, in \inlinecite{Alexander1996} and \inlinecite{Zhao1996}
it was shown that CMEs did not greatly affect the large-scale
coronal structure and the neutral sheet geometry. Using numerical
modeling, \inlinecite{Luhmann1998} have shown that coronal field
lines can be opened without significant changes in the coronal
structure and the neutral line. \inlinecite{Subramanian1999} have
found that although 63\% of CMEs from January 1996 to June 1998
were associated with streamers, the most of CMEs had no effect on
the streamer. The lifetime of the changes in the heliospheric
current sheet (HCS) location caused by CMEs were found to be
significantly less than the lifetime of the HCS structure even
during solar maximum \cite{Zhao1996}.

According to \inlinecite{Ivanov1997} both the properties of CMEs
and their cyclic evolution are closely related to the multipole
component of the global solar magnetic field ($n > 4$),
corresponding to a system of closed magnetic fields on the Sun
with characteristic mean dimensions $D > 40^{\circ}$. CMEs are
caused by the interaction of two large-scale field systems, one of
them (the global field system) determines the location of CMEs and
another (the system of closed magnetic fields) their occurrence
rate \cite{Ivanov1999}. In \inlinecite{Obridko2012} it was
established that CME velocity and occurrence rate depend on the
cyclic variations of the large-scale magnetic fields which
determine active complex evolution and are responsible for the
occurrence of major CMEs. Equatorially trapped Rossby-type waves
were proposed by \inlinecite{Lou2003} as large-scale
quasi-periodic source of the photospheric magnetic field
disturbances, resulting in observed CME periodicities.

While it has been considered that CMEs are a part of the
large-scale magnetic field evolution this connection has not been
investigated in detail. In this paper, the relevance of the CME
occurrence rate and parameters to the GMFS cycle evolution, and
their association with magnetic field oscillations has been
analyzed. The GMFS changes, as a consequence of the excitation of
Rossby waves of different periods and the influence of the changes
in Rossby wave periods on GMFS reorganization and consecutively on
the CME rate and parameters during solar cycles 23 and 24, are
discussed. The comparison with AR parameter cycle evolution is
also studied.

The structure of the paper is as follows. Section \ref{secdata}
describes the used data. In Section \ref{secglobmag} the GMFS
cycle evolution is analyzed. In Section \ref{seccme} CME rate and
parameter cycle changes have been described and the comparison
with that of GMFS and the oscillations in the mean solar magnetic
field, as well as the comparison with AR parameter cycle
evolution, is made. The results are discussed in Section
\ref{secdiscus}.The main conclusions are listed in Section
\ref{seccon}.

\section{Data}   \label{secdata}

The data from the Large Angle and Spectrometric Coronagraph ({\it
LASCO}) on board the Solar and Heliospheric Observatory ({\it
SOHO}) \cite{Brueckner1995} was used. For each CME event, the {\it
SOHO/LASCO} CDAW CME catalogue gives position angle, plane-of-sky
speeds and width, acceleration, mass and energy. Details on the
CME catalog can be found in \inlinecite{Yashiro2004} and
\inlinecite{Gopalswamy2009}. There were data gaps in {\it SOHO}
for June-October 1998 and January-February 1999. In this catalog
CME masses and potential energies may be underestimated by a
maximum of two times, and the kinetic energies by a maximum of
eight times (\opencite{Vourlidas2010}, \citeyear{Vourlidas2011}).
In \inlinecite{Yashiro2008} it was noted that some slow and narrow
CMEs may not be visible when they originate from the solar disc
centre. Due to projection effects, some low-latitude
(high-latitude) CMEs may be misidentified as high-latitude
(low-latitude) CMEs, but no plane-of-sky CME parameter correction
was made, because according to \inlinecite{Howard2008}, in a large
sample of events, plane-of-sky measurements may be suitable for
studies of general trends. Correcting for projection effects is
necessary for those investigations that deal with the properties
of individual CMEs.

To analyze the solar global magnetic field, the data on the mean
solar magnetic field (MMF) and source surface synoptic maps from
the Wilcox Solar Observatory (WSO) were used. For these maps, the
coronal magnetic field is calculated from photospheric fields with
a potential field model with the source surface location at 2.5
solar radii (\opencite{Altschuler1969}; \opencite{Altschuler1975};
\opencite{Altschuler1977}; \opencite{Schatten1969};
\opencite{Hoeksema1986}; \opencite{Hoeksema1988};
\opencite{Wang1992}). Source surface magnetic field data are
consist of $30$ data points in equal steps of sine latitude from
$+70^{\circ}$ to $-70^{\circ}$. Longitude is presented in
$5^{\circ}$ intervals.

For comparison with AR parameter cycle evolution, Space Weather
Prediction Center data were used.

\section{Global magnetic field structure evolution}  \label{secglobmag}

\begin{figure}

 \hspace{-4mm} \includegraphics[width=1.\textwidth]{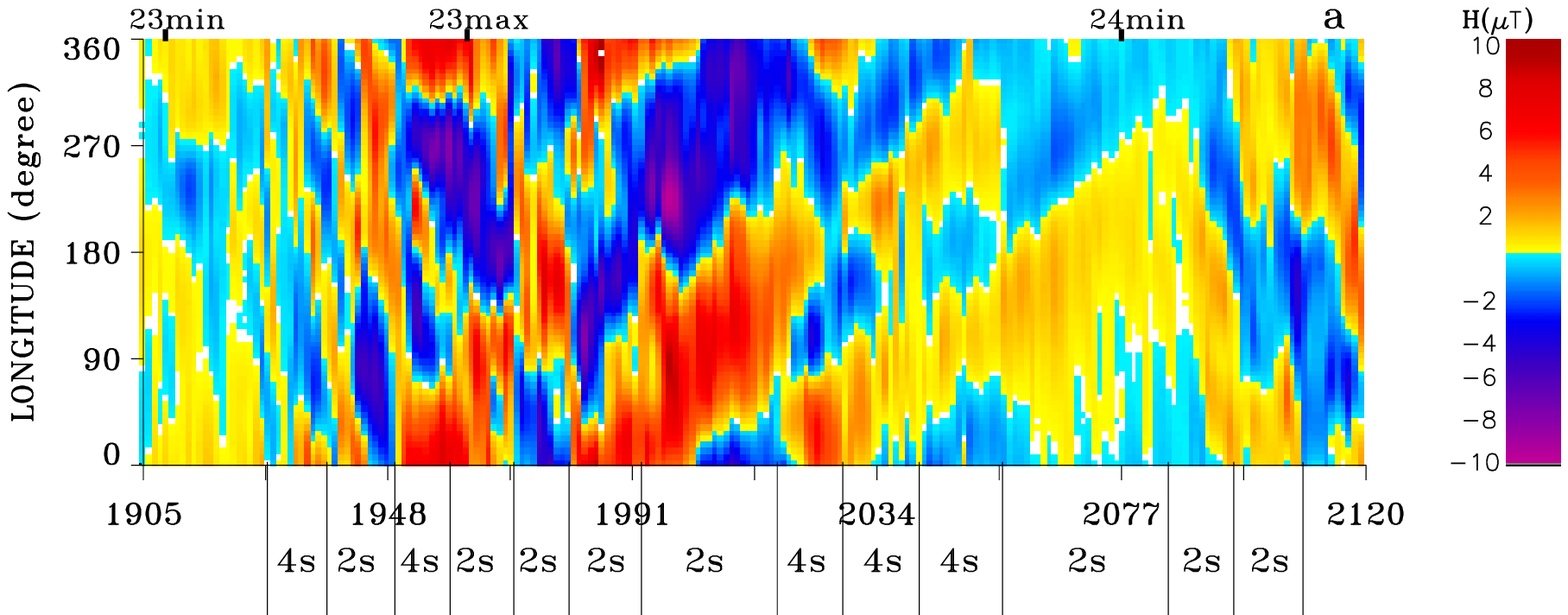}

 \hspace{-4mm} \includegraphics[width=1.\textwidth]{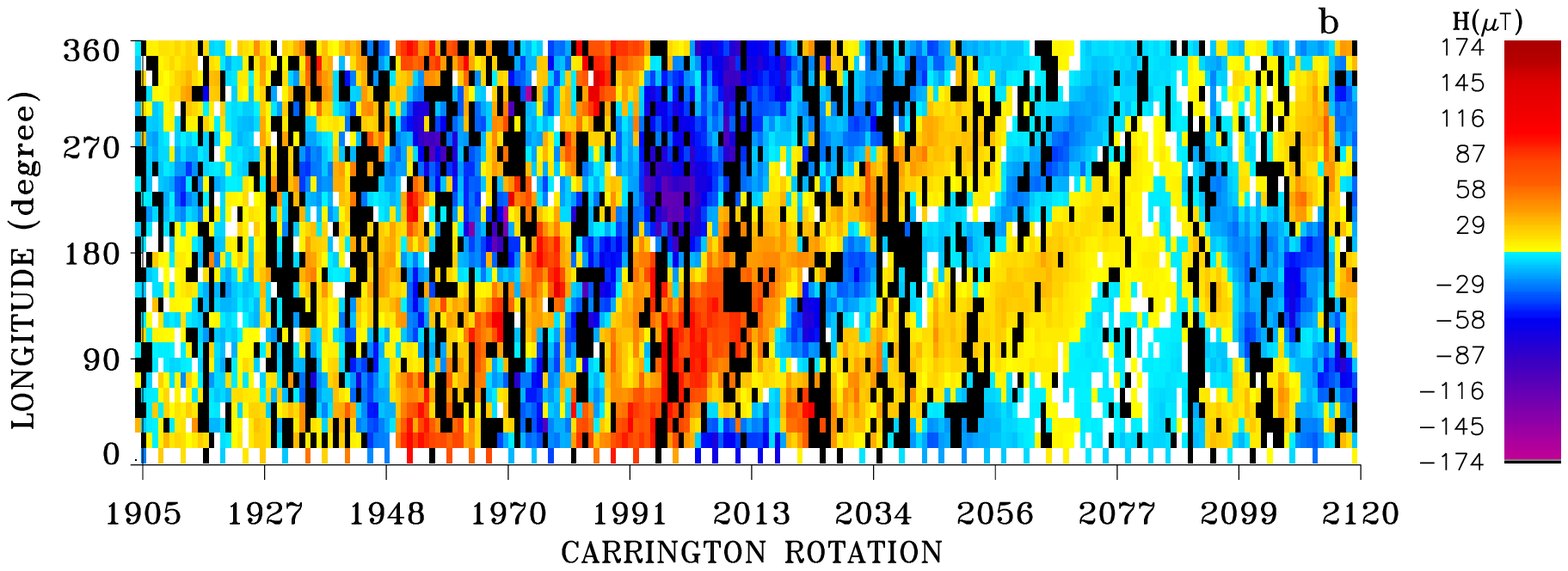}

  \includegraphics[width=1\textwidth]{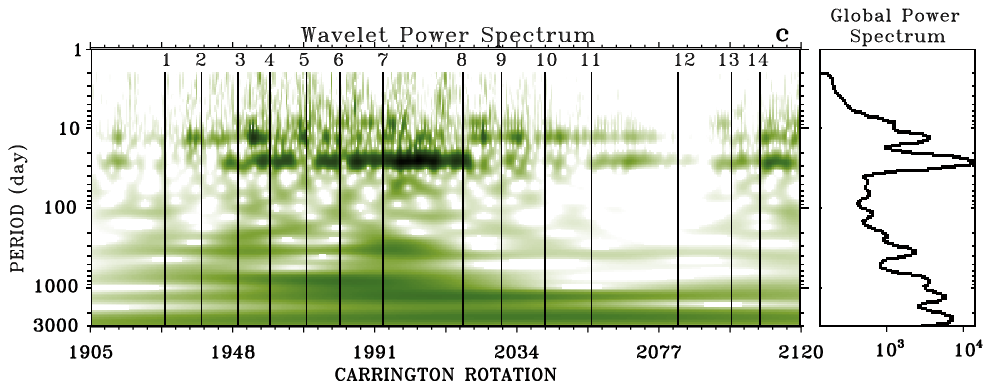}

\caption{(a) The longitudinal diagram of coronal magnetic field
evolution resulting from the PFSS extrapolation at 2.5 solar
radii. (b) The longitudinal diagram of the magnetic field of the
Sun as a star. Yellow-red colors denote positive-polarity magnetic
fields and blue-lilac - negative-polarity fields. Black marks the
missing data. (c) Wavelet power spectrum of the mean solar
magnetic field.}
   \label{londiag}
   \end{figure}

Unfortunately, direct methods to measure the magnetic field in the
solar corona during solar cycles, especially in its quiet regions,
are unavailable to date. Therefore, the WSO calculation results on
coronal magnetic fields in the potential-field approximation with
a standard spherical source surface at 2.5 solar radii (PFSS) for
Carrington rotations (CRs) 1905 - 2119 were employed. The PFSS
model provides remarkably good description of the coronal magnetic
field structure. The longitudinal distribution of
positive-polarity and negative-polarity magnetic fields, resulting
from the PFSS extrapolation at 2.5 solar radii, is displayed in
Figure~\ref{londiag}a. The times of sunspot maximum and minima are
marked at the top of the Figure~\ref{londiag}a. Changes in the
magnetic fields at the source surface reflect those observed over
the same time in the photosphere. The magnetic structure of the
Sun as a star is known to be in good agreement with the PFSS
extrapolation of the coronal magnetic field \cite{Kotov1994}. In
Figure~\ref{londiag}b the longitudinal diagram of the photospheric
magnetic field of the Sun as a star, composed from the data of the
solar MMF, is presented. In Figures~\ref{londiag}a and
\ref{londiag}b colors show the distribution of positive-polarity
(yellow-red) and negative-polarity (blue-lilac) magnetic fields
averaged over latitude for each CR. The brightness at a certain
point is proportional to the magnetic field strength. Black color
marks the missing data.

In the diagrams, Y-axis denotes longitude in degree and X-axis
denotes CR. Each thin vertical bar shows magnetic field
distribution for each CR. When there are two longitudinal
intervals one covered by the positive-polarity magnetic fields and
the other covered by the negative-polarity ones in one CR (along
Y-axis), it means that the two-polarity structure is observed and
when there are two longitudinal intervals covered by
positive-polarity magnetic fields and two longitudinal intervals
covered by negative-polarity magnetic fields in one CR (along
Y-axis), it means that a four-sector structure exists during the
CR. The periods corresponding to the two-sector and four-sector
structures are marked in the Figure~\ref{londiag} as 2s
(two-sector structure) and 4s (four-sector structure). When the
polarity changes from one CR to the next (along the X-axis) in
some longitudinal intervals (Y-axis) it means that the structure
changes its polarity.

The structures in Figures~\ref{londiag}a and \ref{londiag}b
display a good agreement. Thus, the positive-negative-polarity
structure is believed to trace the global solar magnetic field
evolution from the photosphere to the corona. The diagrams show
that positive-polarity and negative-polarity magnetic fields are
not distributed randomly, but rather, form a multi-scale GMFS,
depending on the phase of cycles 23 and 24. Similar structures
were also observed in cycles 21 and 22 (\opencite{Hoeksema1988};
\opencite{Hoeksema1991}; \opencite{Levine1979};
\opencite{Kovalenko1988}).

From Figures~\ref{londiag}a and \ref{londiag}b it is seen that the
lifetime of each sector structure is different during different
cycle phases. During the minimum of the cycle 23 the lifetime of
the observed structures was short approximately 3-5 CRs that was
$\sim80 - 135$ days. During the maximum, the declining phase of
cycle 23, and the minimum of cycle 24 the lifetime of each
structure ranged from $\sim10$ CRs to $\sim1970$ CRs that was from
270 days to $\sim5$ years. A closer look at Figures~\ref{londiag}a
and \ref{londiag}b reveals periods of slow and fast GMFS changes.
There were two fast ($\sim1 - 3$ CRs) redistributions of the GMFS,
covering a considerable part of the solar surface during the
maximum and the beginning of the declining phase of cycle 23.
Two-sector structure with the positive-polarity field domination
at longitudes $330^{\circ}- 360^{\circ}-0^{\circ}-140^{\circ}$ and
the negative field at longitudes $140^{\circ}-330^{\circ}$ was
existing from CR 1959 until CR 1969. The global picture remained
quasi-stable during $\sim300$ days. Then the polarity structure
was reversed during one CR. A new two-sector structure with the
opposite distribution of the positive-polarity and
negative-polarity magnetic fields existed during 10 CRs
($\sim0.75$ year) from CR 1970 until CR 1980, and then in CR 1980
the polarity structure was reversed to the previous distribution
of the positive-polarity and negative-polarity magnetic fields in
one CR. Such redistributions of the GMFS involve the whole Sun.
Magnetic structures were also observed from CR 1905 to 1950 (the
minimum and the rising phase of cycle 23). However, their scale in
longitude and in time was somewhat smaller compared to those of
the maximum and the declining phase.

Why the structures form and disappear and what controls the
regularity in their evolution is not yet understood. Obviously,
the observed distribution and redistribution of magnetic fields
are the consequence of the processes occurring inside the Sun.
\citeauthor{Gilman1969a} (\citeyear{Gilman1969a},
\citeyear{Gilman1969b}) was the first to propose that observed
solar magnetic fields can be the result of Rossby waves in the
Sun's convection zone and photosphere.

The wavelet power spectrum of the daily MMF is shown in
Figure~\ref{londiag}c. To obtain values of the periods
(frequencies) of oscillations contained in the time series it is
necessary to use any method of decomposition of this time series.
Fourier analysis provides the values of the periods (frequencies)
only. The wavelet technique allows obtaining not only values of
periods contained in the analyzed time series, but it also shows
the period's location in time. Thus, we can see when certain
periods appear and disappear. In our case it allows us to compare
the evolutionary changes in the GMFS with periods of oscillations
in the observed magnetic field of the Sun as a star. Morlet
wavelet technique was used. We see oscillations with different
periods during different solar cycle phases. The wave periods
became shorter from $\sim$400 d to 50 d from the minimum to the
maximum of cycle 23, and they grew to the minimum of cycle 24
again. The process was not a smooth one, but it had a form of
individual steps. Each period remained quasi-constant until a new
one appeared. The periods of $\sim$13 d and $\sim$27 d can be
associated with the photospheric magnetic fields and ARs
\cite{Bai1990}. The periods of waves that can be associated with
the GMFS in Figures~\ref{londiag}a and \ref{londiag}b lie in the
range $\sim$50 d - 1000 d (Figures~\ref{londiag}c). The
significant reconfigurations of the GMFS, observed in
Figures~\ref{londiag}a and \ref{londiag}b, are marked by thin
vertical lines. We can see that the changes in the GMFS are
accompanied by the changes in the periods of oscillations that
each wave period coincides in time with a particular GMFS
distribution. Periods from 40 to 600 days do not exist all the
time, but they appear and disappear. They strengthen in the
periods of $\sim200 \div 400$ d from CR 1938 (line 2) to CR 1959
(line 4) coinciding with the formation of the large GMFS. The
appearance of periods $\sim80 \div 110$ d line 3 occur when a
two-sector structure was changed to a four-sector structure, and
these periods disappear at the time marked by the line 4 (CR
1959). The periods became shorter $\sim50 \div 90$ d and $\sim100
\div 200$ d and the GMFS changed from four-sector to two-sector.
The first periods disappear before the line 5 and the second one
at the time marked by line 6. It is interesting to note that the
disappearance of the first periods coincide in time with the
formation of the small negative-polarity feature inside the
positive-polarity pattern at longitudes $\sim40^{\circ} \div
80^{\circ}$. From CR 1970 (line 5) the periods of $\sim50 \div 90$
d disappeared and the waves with periods $\sim80 \div 120$ d
appeared, and the GMFS reversed its polarity. Since CR 1980 (line
6), the periods of $\sim50 \div 100$ d and $\sim120 \div 200$ d
appeared again and the GMFS also reversed its polarity to the
previous state. Line 6 marks the moment when two-sector structure
reverses it's polarity. The periods of $\sim50 \div 70$ d appeared
and their disappearance coincide with formation of extensions from
the positive-polarity patterns at CR $\sim$1980 (line 6). From CR
1993 (line 7) the periods became longer stepwise and the
large-scale two-sector drifting structure was observed. The drift
in the GMFS means the changes in the differential rotation. The
periods of $\sim40 \div 70$ d, appearing at the time marked by
line 7, may coincide with the formation of an extension of
large-scale positive-polarity pattern at longitudes
$\sim160^{\circ} \div 200^{\circ}$. The GMFS from CRs 1980 to 2020
was associated with the same periods as the GMFS of CRs $\sim 1949
- 1970$. Since CR 2017 (line 8), the GMFS became a four-sector
drifting structure, and at the same time we can see step-like
lowering periods, which can be associated with that structure in
the wavelet spectrum. At the rising phase of cycle 24 the new
periods appeared in the wavelet power spectrum, and at the same
time the GMFS changed its drift direction. The intensity of the
cycle 24 period was lower than that of cycle 23.

\begin{figure}
 \vspace{50pt}
 \includegraphics[width=1.\textwidth]{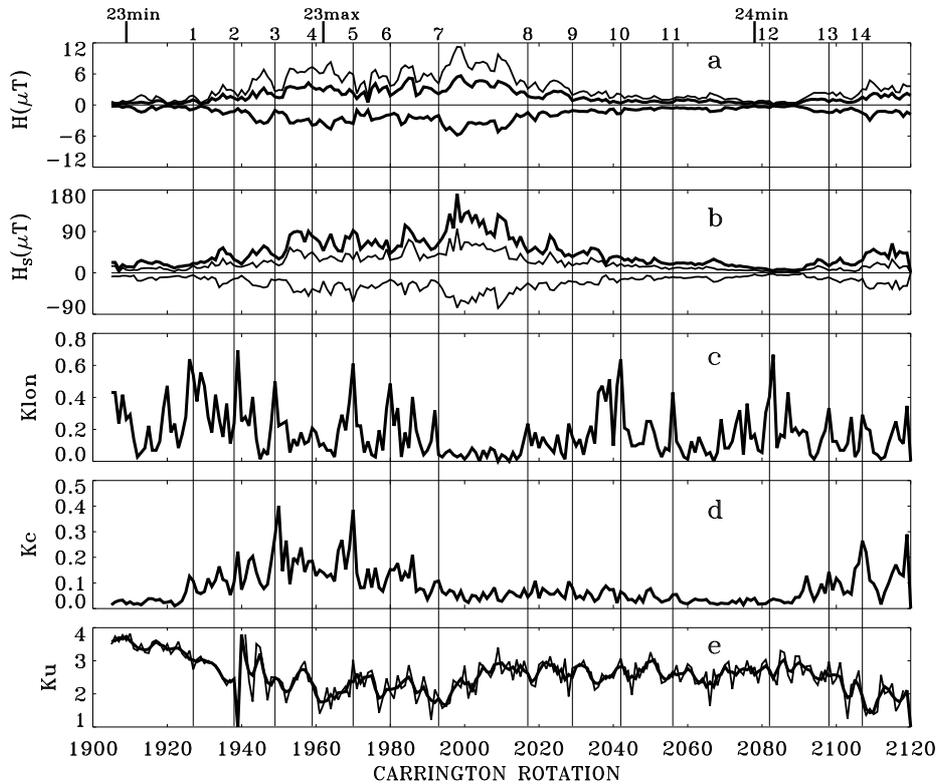}
 \caption{(a) Magnetic field strength from the longitudinal diagram;
          (b) $K_{lon}$;
          (c) $K_{c}$;
          (d) $K_{u}$.
          Thin vertical lines mark the moments of change in the GMFS.}
  \label{globkoef}
\end{figure}

As well as being an interesting phenomenon in its own right, this
behavior of the solar global magnetic field may shed new light on
the observed regularity in CME formation and rate and parameter
evolution during solar cycles. To quantify the changes in the
GMFS, the CR average magnetic field strength of the
positive-polarity and negative-polarity magnetic fields and their
absolute value sum were calculated (Figure~\ref{globkoef}a) using
the longitudinal diagram (Figure~\ref{londiag}a). In
Figure~\ref{londiag}b the CR average magnetic field strength of
the positive-polarity and negative-polarity magnetic fields and
their absolute value sum for the magnetic fields of the Sun as a
star is presented. It is seen that the behavior of magnetic fields
in Figures~\ref{londiag}a and \ref{londiag}b is identical. The
intensity of the magnetic field was low at the minima of cycles 23
and 24. Toward the maxima, the magnetic field strength increased.
The magnetic field strength decreased during the reorganizations
of the GMFS (thin vertical lines in Figures~\ref{globkoef}a and
\ref{globkoef}b). In each pattern, the magnetic field strength did
not grow gradually, but underwent abrupt changes, reflecting
changes in activity within a pattern.

In order to describe the phenomenon of longitudinal structure
changes quantitatively, a series of $K_{lon}$ coefficient was
built \cite{Bilenko2012}. In the longitudinal diagram, magnetic
field polarity in each CR was compared with the successive CR
polarity. The number of points at which the polarity changed was
summed up. Then it was normalized to the total number of
longitudinal points in CR so that $K_{lon}$ takes value from 0 to
1 for each CR. Such normalization allows us to compare the GMFS
change rate at different solar cycle phases.
   \begin{equation}
   K_{lon}=\frac{\sum\limits_{i=0}^{n}{L_{chi}}}{n}
   \end{equation}
\noindent where $L_{chi}$ is the number of longitudes where
polarity was changed between two successive CRs, $n$ - the total
number of longitude points in CR ($n=72$).  Figure~\ref{globkoef}c
presents the $K_{lon}$ coefficient evolution. The amplitude of
$K_{lon}$ decreased up to 0 during the maximum and the decline
phase of cycle 23 when the large quasi-stable structures were
existing. The amplitude of $K_{lon}$ reaches $\sim0.74$ during the
periods of small structures and sharp changes in GMFS. The peaks
in $K_{lon}$ denote the moments of the GMFS reorganization. Thin
vertical lines are drawn through these peaks and the thin vertical
lines in all Figures correspond to these peaks in $K_{lon}$.

To evaluate the total rate of the magnetic field polarity changes,
the coefficient $K_{c}$ was calculated (Figure~\ref{globkoef}d)
from the comparison of polarity in each point in successive source
surface magnetic field maps (WSO). It was normalized to the total
number of points in a map.
 \begin{equation}
    K_{c}=\frac{\sum\limits_{i=0}^k P_{i}}{k},
 \end{equation}
\noindent where $P_i$ is the number of map points where polarity
was changed, $k$ - the total number of points in a map ($k=2160$).
$K_{c}$ reflects the rate of a new magnetic flux emergence.

\section{CME Evolution} \label{seccme}

White light coronagraphs {\it LASCO} have observed nearly 17859
CMEs from 1996 until 2011 (CRs 1905 - 2119). This period covers
almost the whole solar cycle 23 and the beginning of cycle 24.
This large amount of data can help us to improve our knowledge
of CME properties during solar cycles.

\begin{figure}  
 \includegraphics[width=1.\textwidth]{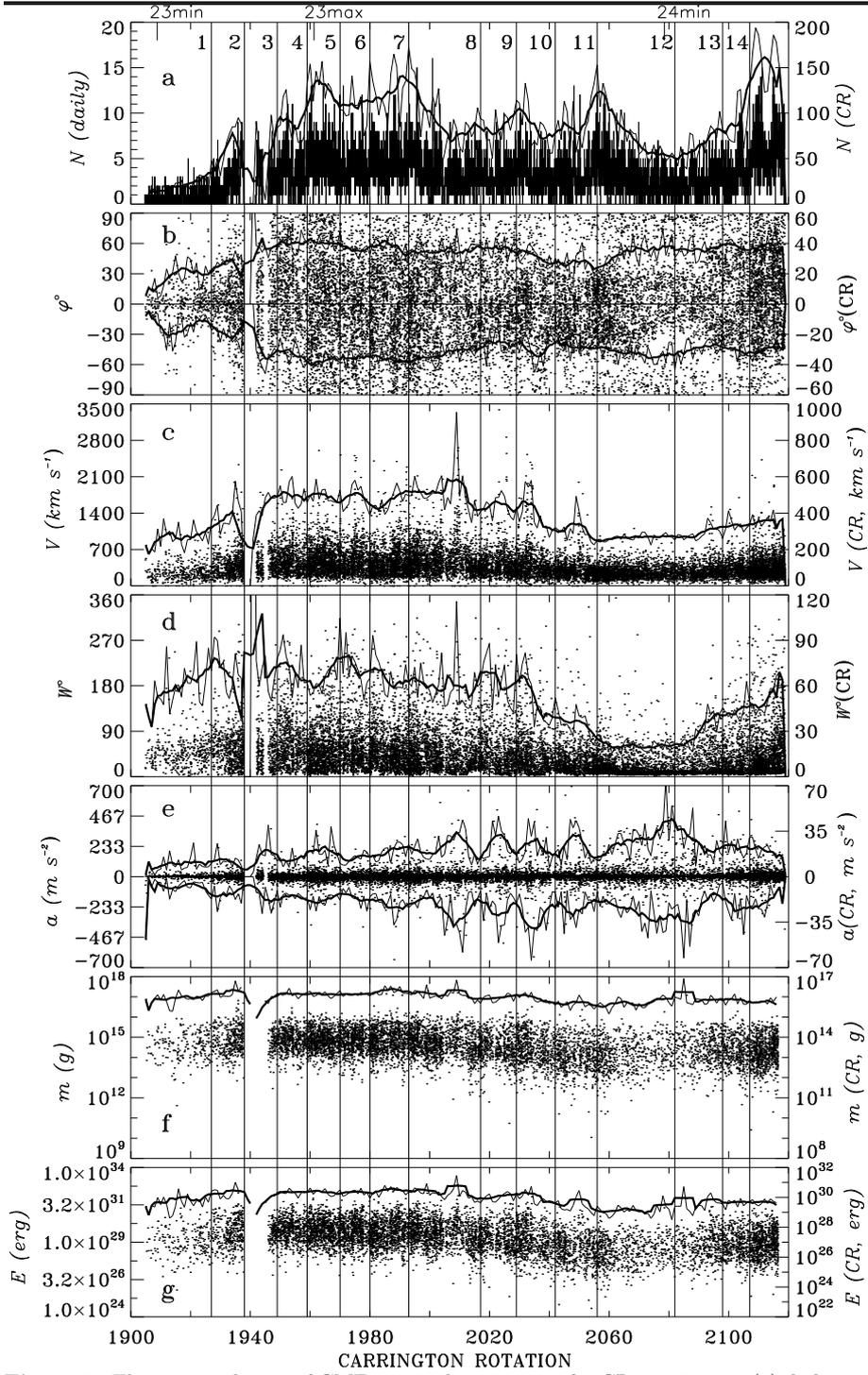}
\caption{The time evolution of CME rate and parameters for CRs
1905-2119. (a) daily rate; (b) the latitudinal distribution of
CMEs; (c) speed; (d) width; (e) acceleration; (f) mass; (g)
energy. {\it Dots} correspond to each CME event. {\it Thin lines}
represent CR averaged data and {\it thick lines} represent 7 CR
averaged data.Thin vertical lines mark the moments of change in
the GMFS.}
 \label{cme}
\end{figure}

Figure~\ref{cme} shows the evolution of CME parameters such as
daily counts of CME events (N), CME latitudinal distribution,
speed (V), width (W), acceleration (a), mass (m) and energy (e) as
a function of time for CRs 1905 - 2119. The position angle (PA) of
each CME was converted to projected heliographic latitude. Dots
represent data for each CME and thin solid lines represent CR
averaged data (the scales are shown on the right y-axis). In order
to filter out high frequency variations in the CME data, CME
parameters were smoothed with 7 CR (approximately half a year)
running mean. The results are shown in Figure~\ref{cme} by thick
lines. Their scales are also shown on the right y-axis. The
significant reconfigurations of the GMFS are marked by thin
vertical lines.

The CME occurrence rate and parameters have a clear dependence on
the phase of a solar cycle. To verify whether CMEs distributed and
occurred randomly in time and space or these variations are not
statistically significant, the test of randomness
(\citeauthor{Wald1943}, \citeyear{Wald1943}) was applied.
 \begin{equation}
     U=\frac{\sqrt{720}\sum\limits_{i=1}^{n-1}\bigl(R_i - \frac{n+1}{2}\bigr)\bigl(R_{i+1} -
     \frac{n+1}{2}\bigr)}{\sqrt{n^2(n+1)(n-3)(5n+6)}},
  \end{equation}
\noindent where R is a time series rank, $n$ is the time series
length. The calculations indicate that all CME, CR, and 7 CR
averaged time series are non-random with $\alpha = 0.95$
confidence level. It means that CME rate and parameter cycle
evolution are the consequence of some regular physical processes.

During solar maxima CMEs erupt at all latitudes
\cite{Hundhausen1993}. Figures~\ref{cme}b shows that CMEs were
distributed over all latitudes including the polar regions in the
maxima of cycles 23 and 24. CME velocity 7 CR averaged amplitudes
remained at the same level during the maximum of cycle 23. There
were only a few outliers in CR averaged data. The period of
oscillations was $\sim$10 CRs. Amplitude of CME width oscillations
diminished from $\sim80^{\circ}$ to $\sim60^{\circ}$. The
oscillation periods was $\sim10 \div 15$ CRs. Acceleration
slightly oscillated with periods changing from $\sim$10 CRs to
$\sim$20 CRs without showing an increase in the amplitude to the
maximum of cycle 23. It can be seen that at the moments of the
reorganization of the GMFS and the changes in MMF oscillations,
described above in the Section \ref{secglobmag} and marked by
vertical lines, the CME rate increased, 7 CR averaged CME
acceleration and velocity decreased. The strength of the magnetic
field decreased that time.

\begin{figure}
  \vspace{-180pt}
  \includegraphics[width=1.\textwidth]{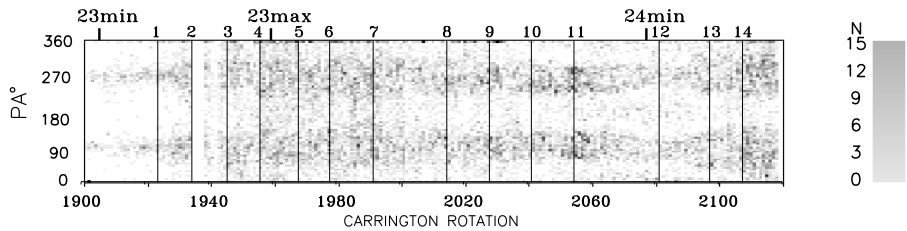}
\caption{The PA distribution of CME occurrence rate calculated for
each CR in PA increments of $5^{\circ}$ for 1996 - 2011 (CRs 1905
- 2119). Thin vertical lines mark the moments of change in the
GMFS.}
 \label{cmelat}
\end{figure}

From Figure~\ref{cme}b we can see that there were periods when the
CME latitudinal distribution was more uniform (CRs: 1958-1962,
1980, 1993, 2017, 2029, 2042, 2056, 2107). It is seen that points
(each point represents an individual CME) show some concentration
to the moments marked by these lines, the moments of the GMFS
reorganization. To analyze the PA (latitudinal) distribution of
CMEs, the occurrence rate was calculated for each CR in PA
increments of $5^{\circ}$. In Figure~\ref{cmelat} the distribution
for each CR versus PA is shown. In order to evaluate the
homogeneity of CME latitudinal distribution, the coefficient of
uniformity $K_u$ was calculated \cite{Frozini1987} from the
distribution in Figure~\ref{cmelat}.

   \begin{equation}
      K_{u}={\frac{1}{\sqrt{n}} \sum_{i=1}^n \Bigl|N_i -
      \frac{i-0.5}{n}\Bigr|},
   \end{equation}
\noindent where $N_i$ is the number of CMEs in each $5^{\circ}$
latitudinal step; $n$ is the number of steps in each CR.
Coefficient $K_u$ describes the latitudinal uniformity of CME
distribution on the solar disc for each CR. Figure~\ref{globkoef}e
presents the $K_{u}$ coefficient evolution. To filter out high
frequency variations, Ku were smoothed with 3 CRs. The growth in
$K_{u}$ means an increase in the inhomogeneity of the latitudinal
distribution of CMEs in a CR. With increasing uniformity in the
CME latitudinal distribution the $K_u$ coefficient decreases. With
the solar activity increase the CME distribution over latitude
became more uniform. Thin vertical lines in Figure~\ref{globkoef}
mark the moments of reconfigurations in the GMFS. The comparison
of Figures~\ref{londiag}a, \ref{londiag}b, and \ref{globkoef}c and
\ref{globkoef}e shows that the moments of the changes in GMFS
(peaks in $K_{lon}$) and $K_u$ decrease (which means the increase
in the CME latitudinal (PA) homogeneity) marked by vertical lines
coincide. This allows us to conclude that the uniformity of CME
latitudinal distribution increases when the GMFS changes. In
Figure~\ref{cmelat} thin vertical lines, marking the same moments,
are shown.  However, the coincidence is not completely accurate,
because, as can be seen from Figures~\ref{londiag}a, and
\ref{londiag}b the reorganization of the GMFS requires at
different longitudes 1-3 CRs. It should be also noted that CMEs
are associated with different solar activity phenomena that can
respond to the reorganization of the GMFS in different ways and
time delay.

\begin{figure}
 \vspace{5pt}
 \includegraphics[width=1.\textwidth]{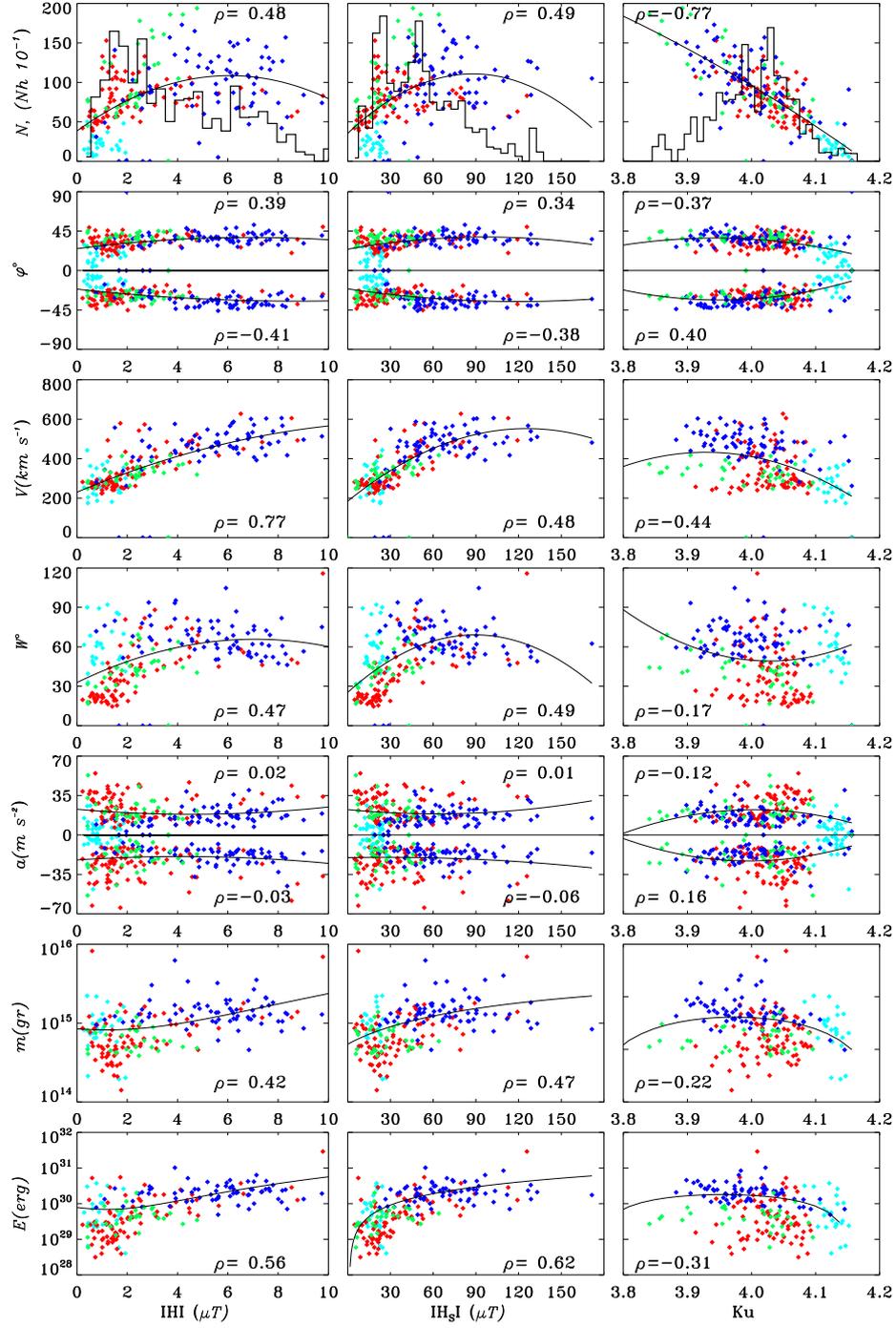}
 \caption{Interrelations between CME parameters and  $|H|$, $|H_{s}|$, and $K_u$.}
  \label{cmehu}
\end{figure}

In Figure~\ref{cmehu} the dependencies of CR averaged CME rate and
parameters on the global magnetic field parameters such as
calculated source surface magnetic field strength $|H|$ and
magnetic field of the Sun as a star $|H_{s}|$ and $K_{u}$ are
summed up. Here, each point represents a CR averaged CME data. All
CME data were divided according to the domination of the zonal or
sectorial structure of the global magnetic field
\cite{Bilenko2012}. Light blue indicates CMEs of the minimum of
cycle 23, which corresponds to the zonal structure domination (CRs
1905-1930). Blue denotes CMEs of the maximum and the beginning of
the decay phase of cycle 23, corresponding to the sector structure
of the solar global magnetic field (CRs 1930-2007). Red denotes
CMEs of the minimum of cycle 24, and the time of the zonal
structure domination (CRs 2007-2087). Green represents the CMEs of
the growing phase of cycle 24, and sector structure domination
(CRs 2087-2119).

Thin lines denote a second-order polynomial fit. All dependencies,
shown in Figure~\ref{cmehu} are non-linear. Therefore Spearman
\cite{Spearman1904} rank correlation coefficient ($\rho$), which
is a non-parametric measure of statistical dependence between two
series, was calculated.

 \begin{equation}
   \rho = 1 - \frac{6 \sum\limits_{i=1}^n (R_{1i} - R_{2i})^2}{n(n^2 - 1)}
 \end{equation}

\noindent where n is the length of time series, $R_{1i}$ and
$R_{2i}$ - the ranks of the compared time series. Correlation
coefficients are shown in each panel. The significance level for
these time series is equal to 0.155. For the most dependencies the
correlation found can be of physical significance with the
exception of acceleration.

In the first row, the histograms are shown. The dependencies of
CME parameters from $|H|$ (the first column) show that they
consist of three groups. The first group includes the majority of
CMEs of the minimum of cycle 23 and the minimum and rising phases
of cycle 24. This group is associated with low magnetic field
strength $|H|< 3 \, \mu T$. The CMEs of the first group are
located near the solar equator, they have low speed from $ 200 \,
km \, s^{-1}$ to $400 \, km \, s^{-1}$, width from $15^{\circ}$ to
$90^{\circ}$, low mass, and energy. The acceleration varies widely
$\sim \pm70 \, km \, s^{-2}$. The second group is associated with
the magnetic field strength from $ 3 \, \mu T$ to $5.5 \, \mu T$.
CMEs have speeds from $250 \, km \,s^{-1}$ to $600 \, km
\,s^{-1}$, width from $40^{\circ}$ to $90^{\circ}$, they have
moderate acceleration, mass, and energy. And the third group is
associated with magnetic field strength greater than $ 5.5 \, \mu
T$. These CMEs have high speed, acceleration, mass, and energy,
but moderate width. The dependencies of CME parameters from
$|H_{s}|$ are presented in the second column. They are also
consist of several groups and have similar distributions and
correlations.

The dependencies of CME parameters from $K_{u}$ are shown in the
third column. The histogtam is peaked around $K_{u}=4.03$. It is
seen, that the maximum uniformity is achieved when the sector
structure of the global magnetic field is dominated (blue and
green, $K_u$ is low). CMEs associated with the zonal global
magnetic field structure distributed less uniformly (light blue
and red, $K_u$ is high).

\begin{figure}
 \includegraphics[width=1.\textwidth]{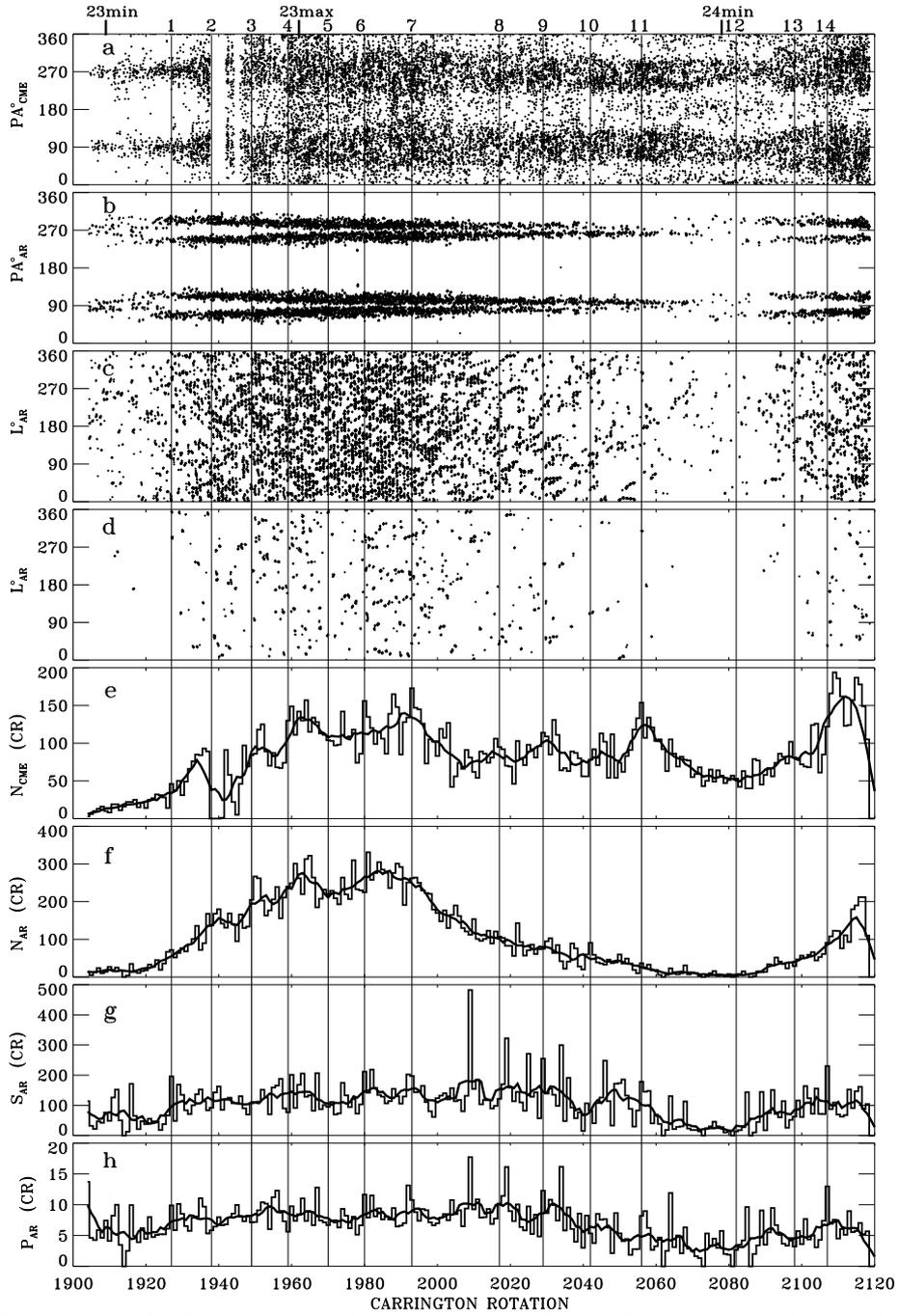}
 \caption{(a) CME position angle;
          (b) AR projected angle;
          (c) AR longitude distribution;
          (d) the longitude distribution of ARs with the area greater than 300 m.v.h.;
          (e) CME occurrence rate;
          (f) AR occurrence rate;
          (g) AR area (m.v.h.);
          (h)  Spot number in each AR.
           {\it Thin lines}, in (e-h), denote CR averaged data and {\it thick lines} correspond to 7 CR
          averaged data. Thin vertical lines mark the moments of change in the GMFS.}
  \label{arlatlon}
\end{figure}

As a rule, most articles on CMEs consider their relation to the AR
local magnetic fields. CMEs occur more commonly where magnetic
fields are stronger, more complex and vary more rapidly
(\opencite{Hildner1976}). But the size of ARs seems to not play an
important role in CME eruptions as sometimes very small ARs are
able to produce CMEs \cite{Schmieder2006}. The fastest CMEs are
known to originate from an instability of AR strong complex
magnetic fields with shear and twist (\opencite{Falconer2002};
\opencite{Gao2011}). Figure~\ref{arlatlon} displays the CME rate
and PA distribution together with the AR distribution and
parameters. Figure~\ref{arlatlon}a shows the PA for all CMEs. For
comparison with CME distribution, the latitude of each AR were
converted to projected angle (Figure~\ref{arlatlon}b). In
Figure~\ref{arlatlon}c, the longitudinal distribution of all ARs,
and in Figure~\ref{arlatlon}d, for those with area greater than
300 millions of visible hemisphere (m.v.h.), are presented. CR
averaged CME occurrence rate ($N_{CME}$) is shown in
Figure~\ref{arlatlon}e. CR averaged AR rate ($N_{AR}$) is shown in
Figure~\ref{arlatlon}f, and the CR averaged area of ARs ($S_{AR}$)
is shown in Figure~\ref{arlatlon}g. The evolution of the number of
spots in each AR ($P_{AR}$) is presented in
Figure~\ref{arlatlon}h. Thin lines in Figures~\ref{arlatlon}e
-\ref{arlatlon}h denote CR averaged data, and thick lines denote 7
CR averaged data. The timing of the changes in the structure of
GMFS are marked by thin vertical lines.

From Figures~\ref{globkoef}, \ref{cme}a, and \ref{arlatlon}e it is
seen that the increase in CME number was not smooth and gradual
during the rising phases, but it had the form of individual bursts
(lines 1, 2, 3, and 13, 14). $K_{lon}$ impulses were rather high,
reflecting the reorganizations of the GMFS in a wide range of
longitudes. The impulses in $K_{lon}$ coincided with the growth in
$K_c$ and an increase in AR number and area. The simultaneous
increase in AR number and area indicates that a new magnetic
fields is formed, in general, by the emergence of a new magnetic
flux forming new ARs \cite{Ballester1999}. The coincidence of CME
and AR impulses suggests that the increase in CME activity was
associated with ARs at that time.

The latitude distribution of CMEs changes during solar cycles
(\opencite{Hundhausen1984}; \opencite{Hundhausen1993};
\opencite{Yashiro2004}; \opencite{Lara2005};  \opencite{Lara2008};
\opencite{Robbrecht2009a}). \inlinecite{Hundhausen1993} noted that
in the beginning of a solar cycle the latitude distribution of
CMEs and ARs is different. The ARs of a new cycle emerged at
latitudes about $\pm30^{\circ}$, and CMEs are concentrated to the
equator. But at the beginning of cycle 24 CMEs were widely spread
over the solar latitude.

From Figures~\ref{arlatlon}e and \ref{arlatlon}f it can be
inferred that just like the AR, the CME occurrence rate rose
during the growing phase and decayed after solar maximum. The
decay was not a smooth exponentially decaying process, but it had
a number of peaks that were comparable in magnitude to the values
of CME parameters at the maximum of cycle 23. In
Figures~\ref{cme}b, \ref{arlatlon}a the general latitudinal drift
of low-latitude CMEs towards the equator during the declining
phase is clearly distinguished. It most likely reflects the
changes in the latitudinal distribution of the CMEs associated
with ARs. But the decay rate of CMEs was lower than that of ARs.
\inlinecite{Robbrecht2009a} have proposed that some CMEs
originated from non-sunspot regions at that time. According to
\inlinecite{Obridko2012} the CMEs can be associated with the giant
cells and coronal holes.

It is well known that there are two peaks in Wolf number separated
by Gnevychev gap (the dip in the solar activity). Secondary solar
activity peak occurred two to three years after the main maximum
(\opencite{Gnevyshev1963}, \citeyear{Gnevyshev1967}). In cycle 23
the first peak was in April 2000 (CR 1962, $ W=120.8 $) and the
second peak occurred in November 2001 (CR 1983, $ W=115.5 $). CME
rate, Figures~\ref{cme}a and \ref{arlatlon}e, also shows two peaks
and the gap. The gap in CME evolution was also retrieved in the
CACTus data \cite{Robbrecht2009a}. According to
\inlinecite{Robbrecht2009a} and \inlinecite{Ramesh2010}, the CME
second peak shows a delay of 6 to 12 months with respect to the
sunspot index. Our result shows the delay for the second peak was
equal to 10 CRs.

From Figures~\ref{cme} and \ref{arlatlon} we can see that during
each peak, the growth of CME and AR number was in the form of
individual impulses. But the rate of CMEs closely followed the ARs
only during the rising phase and the first peak. They were
differed greatly during the second peak and the declining phase.
Comparison with Figure~\ref{londiag} shows that the first peak
occurred when the large two-sector structure appeared. From
Figures~\ref{londiag}a and \ref{londiag}b, it is seen that the
Gnevychev gap coincides with the change in GMFS and the periods of
oscillations. The periods of $\sim40 \div 80$ d and $\sim100 \div
200$ d weakened and disappeared and the periods of of $\sim40 \div
80$ appeared. A quasi-stable two-sector structure existed from CR
1970 to CR 1980 (lines 5, 6). The structure coincided with the gap
in CME rate. The number of CMEs diminished, they had on average
higher velocity, lower acceleration and rather narrow width. Since
CR 1980 (line 6) the periodicities with periods of $\sim 50\div100
$ d appeared again, the periods of $\sim 100 \div 200 $ d began to
grow and the GMFS reversed the polarity. The CME second peak (line
7, CR 1993) coincided with the changes in GMFS at longitudes
$90^{\circ} - 180^{\circ}$ with the disappearance of periods of
$\sim150\div200$ d and appearance of periods $\sim100\div120$ d
and $\sim40\div80$ d. There was no large new magnetic flux
emergence ($K_c \sim0.1$). The uniformity of CME latitudinal
distribution was very high. The rate of CMEs during the second
peak was higher than that during the first peak. The scale of the
rearrangement of the GMFS was practically the same. Moreover, the
positive-polarity and negative-polarity magnetic fields of the new
structure appeared at the same longitudes as those of the
structure existing at the time of the first peak, and the periods
of oscillations were the same. The GMFS reorganization during the
second CME peak covered a larger longitudinal interval compared to
the first one ($K_{lon}$ was equal to $\sim$0.5). The magnetic
field strength was also higher. The uniformity of CME latitudinal
distribution increased (Figure~\ref{londiag}e). In the
longitudinal distribution of ARs (Figure~\ref{arlatlon}c,
\ref{arlatlon}d) it is seen that the majority of ARs were observed
from CR 1940 to 2000, when the long-lived two-sector structures
with a high magnetic field strength were existing. The comparison
of CME and AR PA distributions, Figure~\ref{arlatlon}a and
\ref{arlatlon}b, shows that they are similar for CME population
concentrated to the AR latitudes. But equatorial CMEs spread over
a much wider region than ARs. Some of such CMEs can be the result
of eruption of cross-equatorial arcs connecting ARs located in the
North and in the South hemispheres \cite{Lara2008}. Some CME
source regions may be close to one of the ARs from
cross-equatorial arcs and suffer a strong deflection toward the
equator \cite{Lara2008}. Some high-latitude CMEs could be the
projections of processes at the latitudes of ARs, or may be the
result of non-radial propagation of erupted filaments caused by
the global magnetic field configuration \cite{Filippov2002}.

During the declining phase, the sharp extensions in AR area were
seen in Figure~\ref{arlatlon}g. But the number of ARs did not
increase. According to \inlinecite{Ballester1999}, if the increase
in the area of AR is not accompanied by an increase in the number
of ARs, it means, that there is a new magnetic flux emerging in
already existing ARs. In Figure~\ref{arlatlon}h the increase in
spot numbers in each AR, coinciding with the AR area impulses, is
observed. Therefore, the complexity of ARs increased. Such ARs are
known to be the sources of flares and, obviously, eruptive events.
This may explains the increase in high velocity CMEs during the
decay phase. High velocity, narrow CMEs with high acceleration
were probably the consequence of the processes occurring in ARs
(CME parameters between the lines 8-11). The comparison of
Figures~\ref{cme} and \ref{arlatlon} shows that the oscillations
in CME parameters, observed during the declining phase, were the
result of the alternation of two processes. The first one was the
emergence of a new magnetic flux coinciding with the AR area and
it's complicity growth. The CMEs associated with that process
(between vertical lines 8-11) had, on average, higher velocity,
lower width and higher acceleration. The flux emergence in ARs
could be the source of solar flares. CMEs associated with flares
have higher V. The second process was associated with the GMFS
reorganizations. The CMEs associated with the GMFS reorganization,
marked by vertical lines (8-11), were characterized by low
velocity, low acceleration, yet higher width.

\begin{figure}
 \vspace{5pt}
 \includegraphics[width=1.\textwidth]{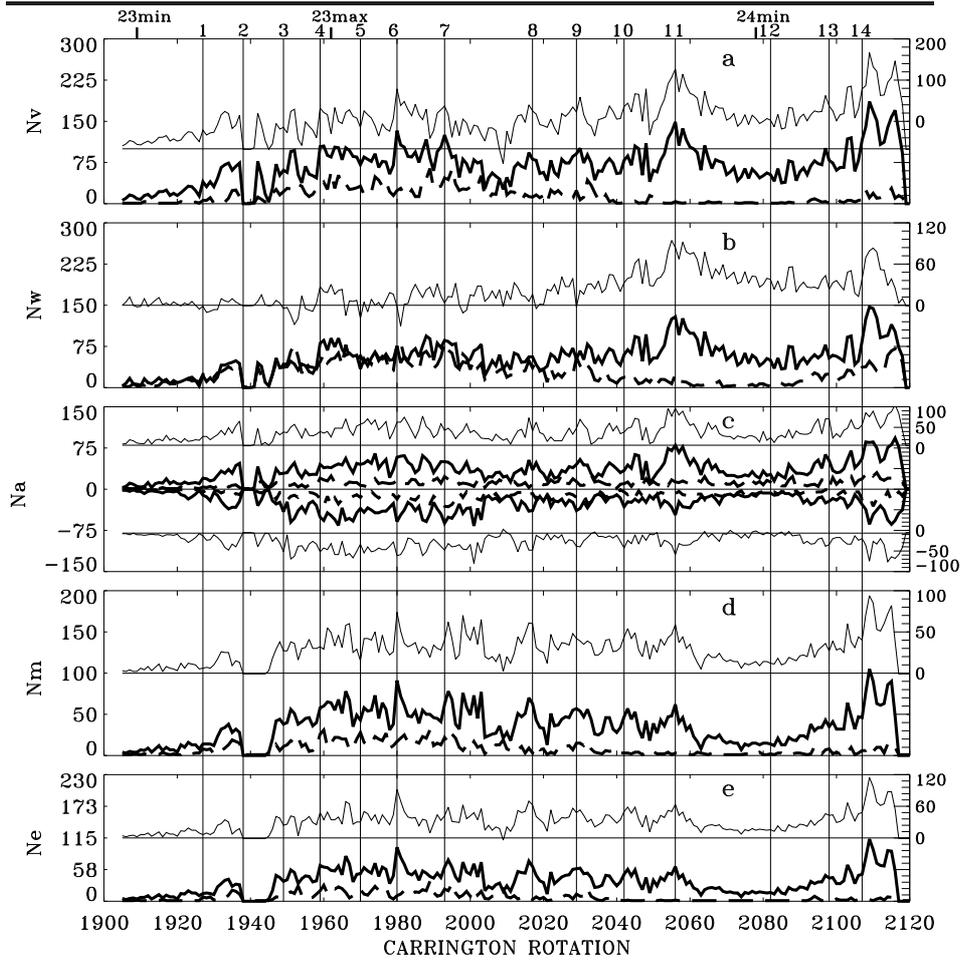}
 \caption{(a) {\it dashed line} the number of CMEs with $V>600 \, km \, s^{-1}$ and {\it solid line} those with $V\le600 \, km \, s^{-1}$;
  (b) {\it dashed line} the number of CMEs with $W>50^{\circ}$ and {\it solid line} those with  $W\le50^{\circ}$;
  (c) {\it dashed line} the number of CMEs with $a>\pm20 \, m \, s^{-2}$ and {\it solid line} those with  $a\le\pm20 \, m \, s^{-2}$;
  (d) {\it dashed line} the number of CMEs with $m>1.7 10^{15} \, g$ and {\it solid line} those with  $m\le1.7 10^{15} \, g$;
  (e) {\it dashed line} the number of CMEs with $e>2. 10^{30} \, erg$ and {\it solid line} those with  $e\le2. 10^{30} \, erg$.
      {\it thin lines} the difference between the number of CMEs with lower and higher than a corresponding limit for each
      parameter. Thin vertical lines mark the moments of change in the GMFS.}
  \label{cmeparam}
\end{figure}

As has been shown above, the number of CMEs during the CRs of the
GMFS reorganization increased and the parameters of the CMEs were
different from those during the periods of quasi-stable GMFS. In
Figure~\ref{cmeparam} the number of CMEs, depending on their
parameters, is shown. The changes in the structure of the GMFS are
marked by thin vertical lines.

At the times of the GMFS reorganizations, market by vertical
lines, the relative number of weak, low-speed, low-accelerated
CMEs with low mass and energy, increase greatly compared to power
events, except the lines 1, 2, 5, and 12. Line 1 corresponds to
the beginning of cycle 23. At that time, no significant events
were observed. Line 5 corresponds to the maximum of cycle 23. But,
during solar maxima the number of small, faint events is
underestimated due to the occulting effect of power CMEs. Line 12
corresponds to the deep minimum of activity, when there are very
few events related to the CMEs. The difference (thin lines) in the
number of CMEs with parameters lower the limits and that greater
the limits increased during the GMFS reorganizations. It suggests
that, in general, fast variations in GMFS make the conditions
favorable for weak, low-mass, low-energy highly accelerated CMEs.
The number of CMEs with $V>600 $ shows local increases between the
lines (between the GMFS reorganizations). The number of CMEs with
the width greater than $50^{\circ}$ and that of CMEs with the
width less than $50^{\circ}$ was almost the same during the
maximum. The decrease in both of them coincided with the Gnevishev
gap and they peaked when large-scale structures were existing. But
they were very different in the decay phase. The number of CMEs
with $W < 50^{\circ}$  and the difference between the number of
CMEs  $W < 50^{\circ}$ and that  $W > 50^{\circ}$ increased
greatly at the moments of the GMFS reorganizations during the
decay phase (line 11).

\section{On the possible relation of Rossby waves and CMEs}   \label{secrosbywave}

Gilman proposed that observed solar magnetic fields can be the
result of Rossby waves generated around the thin magnetized layer
at the bottom of the convection zone (\opencite{Gilman1969a};
\citeyear{Gilman1969b}; \citeyear{Gilman1999}). Rossby waves
belong to a subset of global waves that can exist in a fluid layer
on the surface of a rotating sphere. On the Sun Rossby waves are
strong during the maxima phases and have periods greater than
solar sideral rotation period (\opencite{Lou2000};
\opencite{Zaqarashvili2010a}; \opencite{Zaqarashvili2010b}). In
\citeauthor{Tikhomolov1995} (\citeyear{Tikhomolov1995},
\citeyear{Tikhomolov1996}) Rossby vortices were considered to
explain the observed GMFS. It was proposed that Rossby vortices
were excited within a thin layer beneath the convection zone. They
are a result of heating from the solar interior and the
deformation of the convection zone lower boundary. According to
\citeauthor{Zaqarashvili2010a}, (\citeyear{Zaqarashvili2010a},
\citeyear{Zaqarashvili2010b}) the periodicity of 155-160 days and
$\sim$2 years, observed in different solar activity indices, can
be connected to the dynamics of magnetic Rossby waves in the solar
tachocline, since in the layer they are unstable due to the joint
effect of the toroidal magnetic field strength and latitudinal
differential rotation. It was also proposed that equatorially
trapped Rossby-type waves might modulate solar flares, ARs
\cite{Lou2000} and CME \cite{Lou2003} activity. But in GALLEX
(GALLium Experiment) data the periodicities  of 52 d, 78, d and
154 d were also revealed (\opencite{Sturrock1997}; 1999). The
analysis has shown that the solar neutrino flux exhibits a
periodic variation that may be attributed to rotational modulation
occurring deep in the solar interior, either in the tachocline or
in the radiative zone \cite{Sturrock1997}. These periodicities
probably result from Rossby-type waves occurring in the solar
interior. It means that Rossby waves are generated deep in the
solar interior such as the base of the convection zone, rather
than at the photosphere. In \citeauthor{Kuhn2000}
\citeyear{Kuhn2000} the observation evidence, confirming the
existence of Rossby waves in the photospheric magnetic field,
using observations of the Michelson Doppler Imager (SOHO), is
presented.

\begin{figure}
 \hspace{-2mm}\vspace{1mm}\includegraphics[width=1.\textwidth]{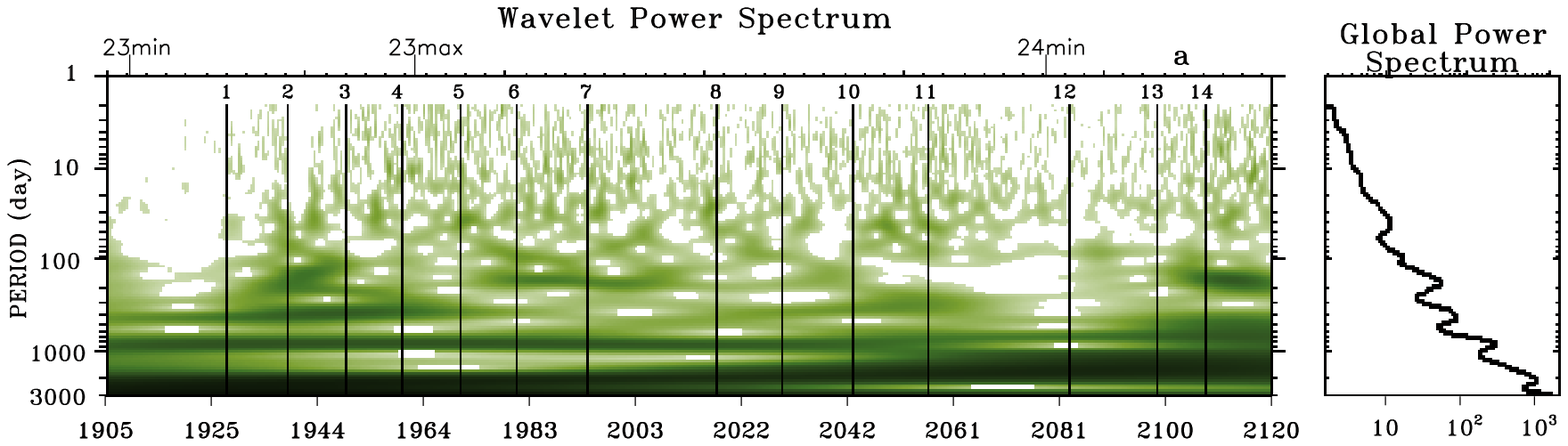}

 \vspace{5mm}
 \hspace{-2mm}\includegraphics[width=1\textwidth]{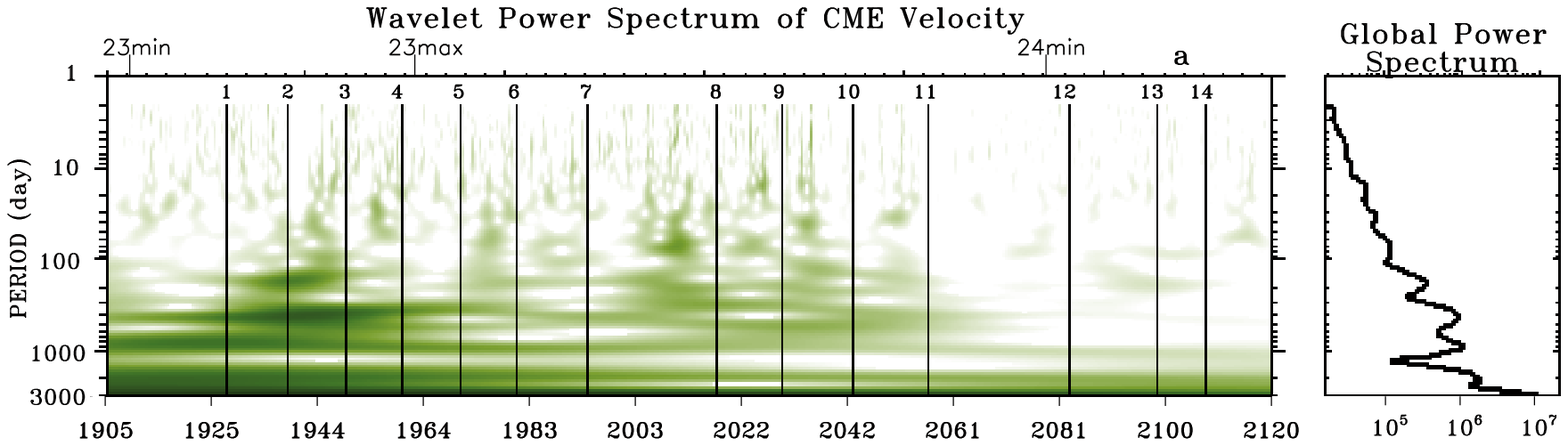}

 \vspace{5mm}
 \hspace{-2mm}\includegraphics[width=1\textwidth]{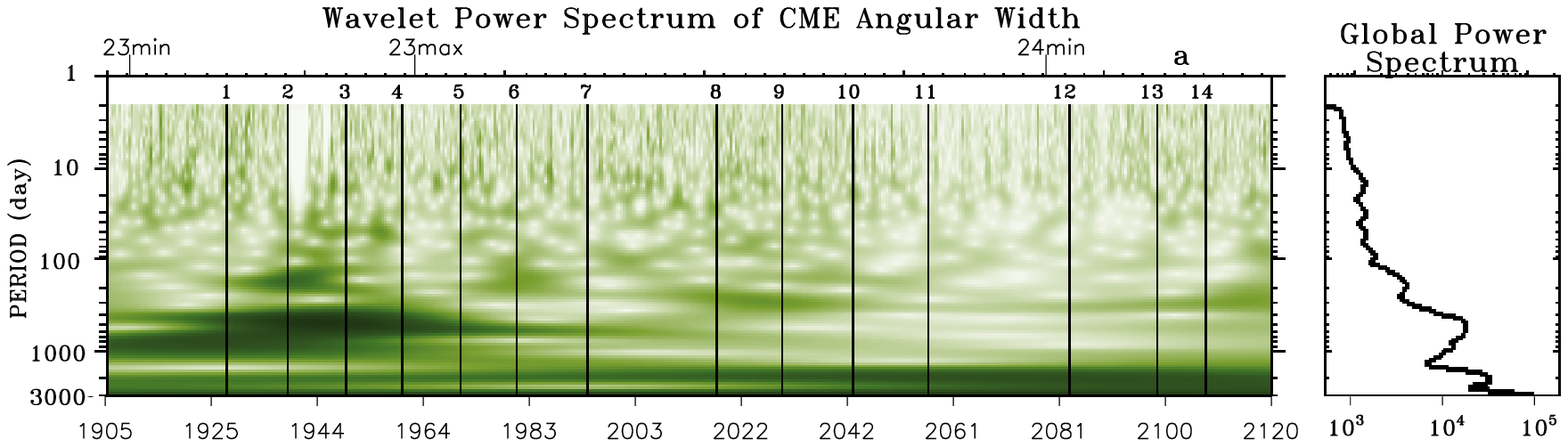}

 \vspace{5mm}
 \hspace{-2mm}\includegraphics[width=1\textwidth]{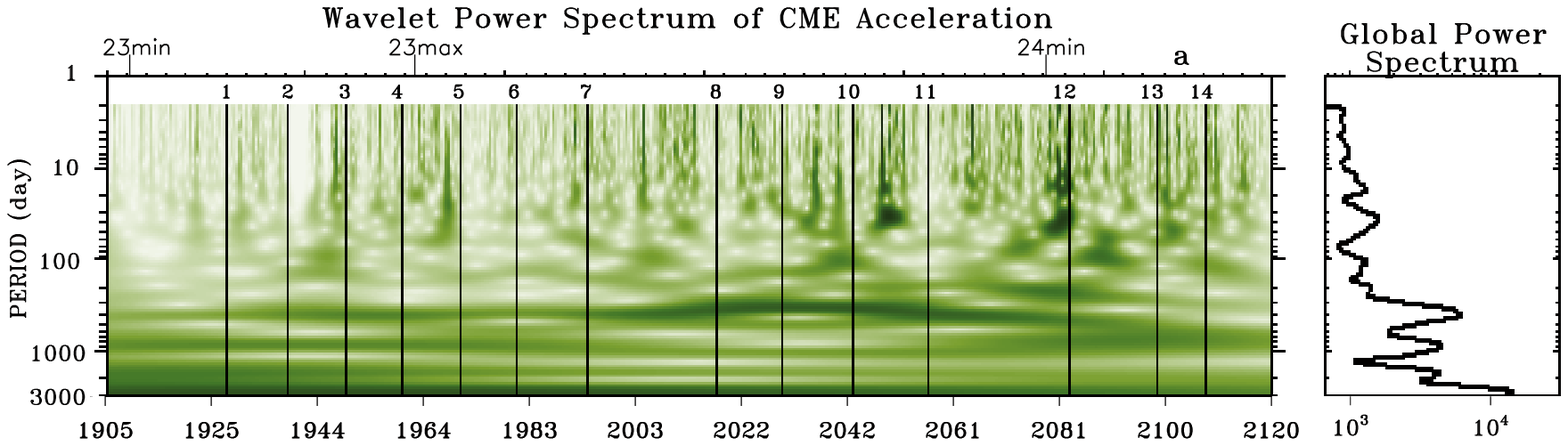}

 \vspace{5mm}
 \hspace{-2mm}\includegraphics[width=1\textwidth]{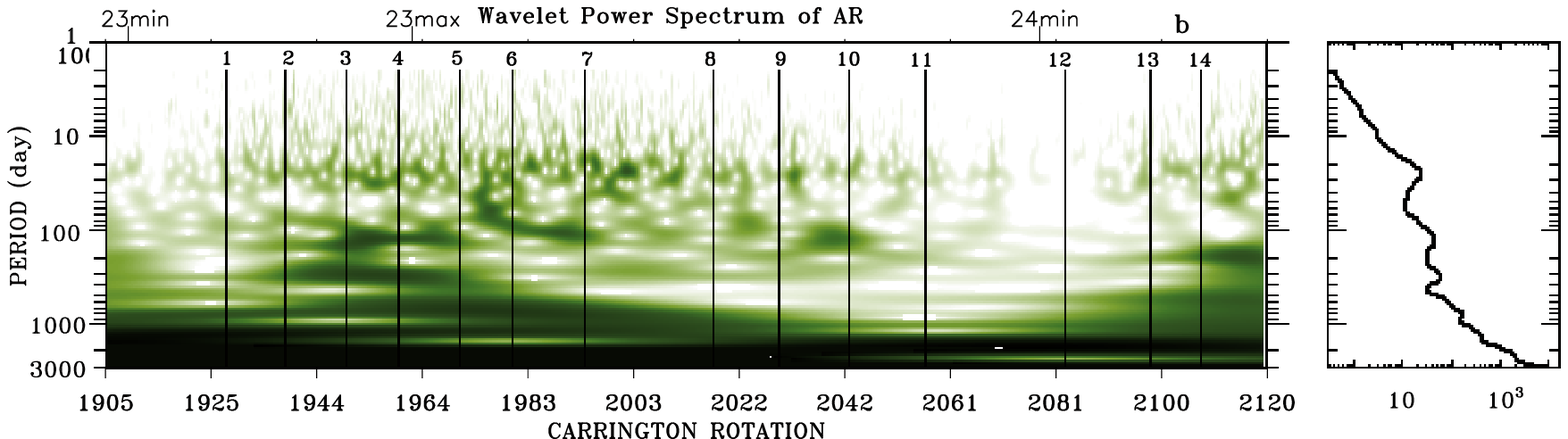}
\caption{The wavelet power spectra of (a) CME daily rate;
   (b) CME velocity; (c) CME angular width;
   (d) CME positive acceleration;
   (e) AR.
   Thin vertical lines mark the moments of change in the GMFS.}
   \label{cmewave}
\end{figure}

\begin{figure}
 \vspace{5pt}
 \includegraphics[width=1.\textwidth]{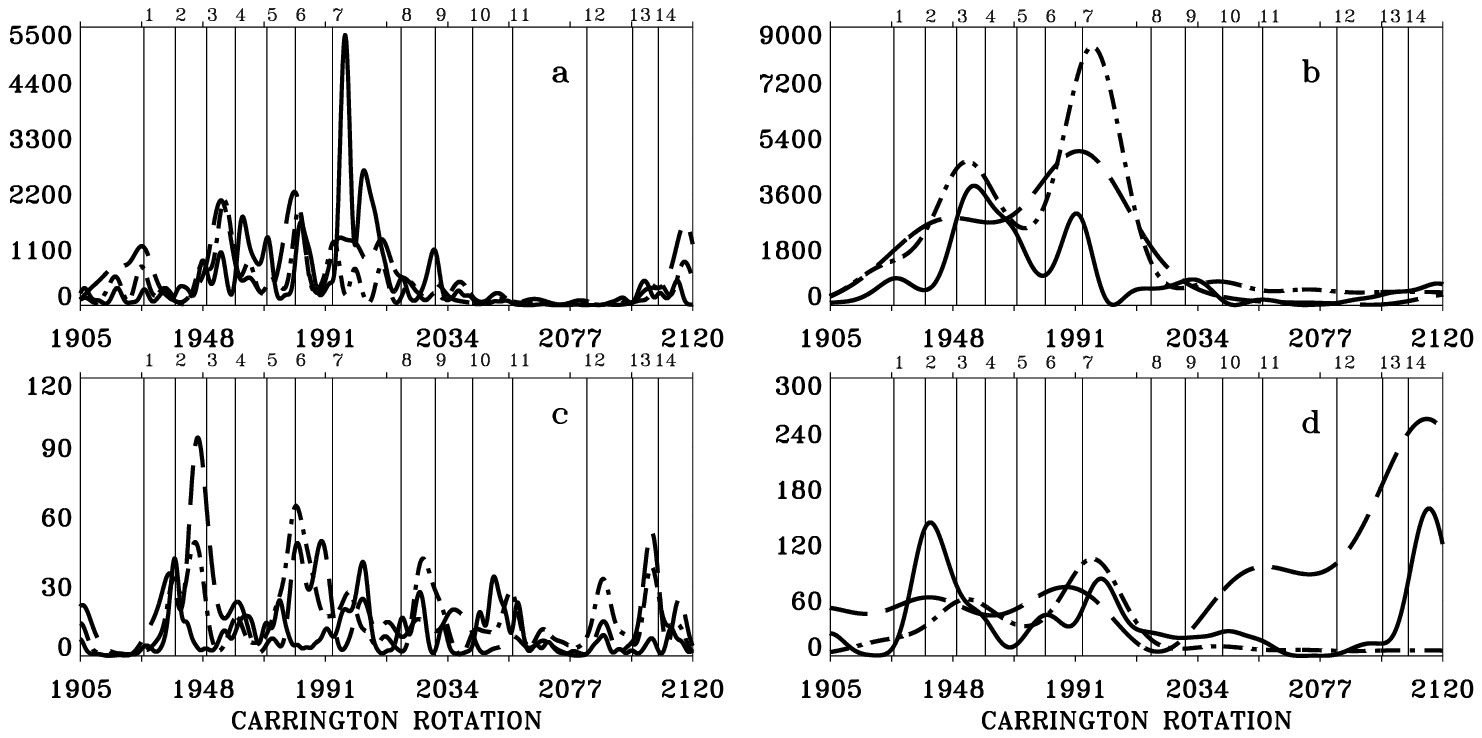}
 \caption{(a) and (b) periods from the wavelet decomposition of MF.
          (c) and (d) periods from the wavelet decomposition of CME number.
  In (a) and (c) {\it solid line} P=50-70d; {\it dash dot line} P=90-100d; {\it long dashes} P=100-120d.
  In (b) and (d) {\it solid line} P=200-220d; {\it dash dot line} P=300-400d; {\it long dashes} P=600-700d.
  Thin vertical lines mark the moments of change in the GMFS.}
  \label{waveper}
\end{figure}

From Figure~\ref{cme} it is seen that the oscillations in CME rate
and parameters did not coincide for different CME parameters and
were different at different solar cycle phases. The strongest
oscillations were observed during the decay phase. The amplitude
of oscillations was about $400 \, km \, s^{-1}$ for individual CME
velocity and $\sim200 \, km \, s^{-1}$ for 7 CR averaged data
(Figure~\ref{cme}).  The intense spike (CR 2009) in the CR
averaged CME velocity and width was due to the Halloween 2003
storm \cite{Gopalswamy2005}. The oscillations were most pronounced
in the CME acceleration. The positive and negative accelerated
CMEs had almost the same amplitudes and they varied synchronously
(Figure~\ref{cme}e). The CME parameters and daily rates have been
analyzed using a Morlet wavelet technique to look for the presence
of the periods and the temporal evolution of these periods
(Figures~\ref{cmewave}a-d) and Figure~\ref{waveper}). Wavelet
power spectrum of AR daily rate is also shown in
Figure~\ref{cmewave}e. Wavelet spectra for periods greater than
$\sim3000$ d are not reliable, as these periods are not much less
than the total data length of 5475 d (edge effects). The strong
peaks around $\sim27$ d were defined in the MMF
(Figure~\ref{londiag}c) and in AR spectra, but they do not present
neither in CME rate nor in CME parameter wavelet spectra
(Figure~\ref{cmewave}a-d). The first peak was around $\sim30-40$ d
in CME spectra. In CME wavelet spectra periods greater than 100 d
dominate. Periods of $300\div400$ d were present in MMF and in
CME. They were absent in AR since CR 2017. There is also a match
around a periodicity of $\sim100 - 150$ d in both MMF and CME
during CRs 1993 - 2017 (lines 7-8). These periodicities were also
not present in AR. MMF and CME have also comparable lowering
frequency oscillations from CR 2017 to CR 2056. This may be
interpreted as large-scale magnetic field driven periodicities in
CME data. It can be seen that there is a temporal coincidence
between the periods corresponding to MMF and CME spectra and again
when the periodicity appeared/disappeared in MMF they also
appeared/disappeared in CME (Figures~\ref{cmewave}a-d) and
Figure~\ref{waveper}). Comparison of Figure~\ref{cme} with
Figure~\ref{londiag} and \ref{cmewave} shows that the peaks in CME
number coincided with the formation of new periods and the GMFS
reorganizations. This comparison makes a strong case for a
relationship between the appearance of the periodicity in CMEs and
the changes in the GMFS. A time and frequency coincidence between
both periodicities suggests the existence of a casual link between
them. The time coincidence in the waves appearing/disappearing is
thought to be very important and seems to confirm the existence of
a casual link between them. Some periods in CME spectrum are
associated with ARs, especially during the maxima of cycle 23. The
periods of $50\div100$ d coincided in CME and AR from CR 1970 to
CR 1993. The fluctuations in CME rate and velocity during the
maximum and the declining phase of cycle 23 were also noticed by
\citeauthor{Ivanov2001} \shortcite{Ivanov2001},
\citeauthor{Lou2003} \shortcite{Lou2003},
\citeauthor{Gerontidou2010} \shortcite{Gerontidou2010}. In
\inlinecite{Lou2003} it was proposed that oscillations with
periods 51, 76, 128, and 153 days in CME daily rate would be
consistent with the presence of large-scale equatorially trapped
Rossby-type waves.

At the beginning of cycle 23, since CR 1949 (line 3) a long-lived
two-sector structure with high magnetic field strength was
created. The structure was the result of the large amount of a new
magnetic flux emergence in a wide longitudinal range
(Figures~\ref{globkoef}c, \ref{globkoef}d). This could be a
consequence of the appearance of the oscillation within the period
from 100 d to 200 d (Figures~\ref{londiag}c, \ref{cmewave}). The
changes of the periods resulted in the changes in the GMFS and in
CME rate growth. Between CRs 1949 - 1959 (between lines 3 and 4) a
short-lived unstable structure appeared at longitudes
$\sim120^{\circ} - 250^{\circ}$ and, simultaneously, the
oscillations with period of $\sim50 - 90$ d was observed. The
configuration of the GMFS changed from a two-sector to a
four-sector and the rate of CMEs diminished. The CMEs associated
with that short-lived four-sector structure were, on average, wide
and had low acceleration (Figure~\ref{cme}).

Lines 8 - 11 (CRs 2017, 2029, 2042, 2056) were associated with
impulses in $K_{lon}$ (GMFS reorganization), but $K_c$ (new
magnetic field emergence) was low. From, the line 8 four-sector
structure was formed and large-scale drifting structures were
observed. The step-like decrease in the wave periods was observed
in MMF and CME (Figure~\ref{londiag}c, \ref{cmewave}). The
decrease was in the form of separate steps and transition to the
each next step was accompanied by the increase in CME number with
low speed, width and acceleration. Line 8 (CR 2017) also
corresponds to the appearance of a new GMFS at longitudes
$0^{\circ}-80^{\circ}$. GMFS changes were not accompanied by a
substantial new magnetic flux emergence. The strength of the
magnetic field was also low. Since line 9 (CR 2029) the two-sector
structure became the four-sector one. The rate of CMEs increased
at that moment. On average, the CMEs were faint. During CRs
2035-2045 (line 10) the two-sector structure was restored for a
short time. That moment was also characterized by an increase in
CMEs with low parameters. The sharp decrease in the rate of CMEs
since 2003 ($\sim$ CR 2000) coincided with the decrease of the
sector structure contribution \cite{Bilenko2012}. A significant
reduction in CME number was also observed from that time.  On
average, the CMEs associated with those periodicities had low
speed and acceleration, but they were rather wide.

Since CR 2056 (line 11) the waves at the range from $\sim40 \div
300$ d disappeared (Figure~\ref{londiag}c). The GMFS changed from
four-sector to two-sector. Magnetic fields changed their polarity
from positive to negative in the longitudes $180^{\circ} -
330^{\circ}$ during one CR. But, as it happened during the decay
phase, there were no large powerful ARs at that time
(Figure~\ref{arlatlon}). There were only very weak new magnetic
flux emergence impulses (Figure~\ref{globkoef}d). The intensity of
the magnetic field was also low. The rate of CMEs was equal to
that of the solar maximum peaks, but the CMEs were faint. They had
low speed, small width and low acceleration.
Figure~\ref{globkoef}e shows that the uniformity of CME latitude
distribution increased greatly at that time.

\section{Discussion}   \label{secdiscus}

According to the previous studies for cycles 21, 22, and 23, CMEs
are concentrated to the solar equator in the range $\pm30^{\circ}$
during solar minima, and their distribution is almost normal. This
is consistent with our findings for cycle 23, however, the
behavior of CMEs at the minimum of cycle 24 was very different.
They were observed at all latitudes. The difference in CME
latitudinal distribution may be explained if we take into
consideration the difference of the GMFS in the minima of cycles
23 and 24. Near cycle 23 minimum, CRs 1900 - 1930, the coronal
field was dipolar, the magnetic field strength was low
(Figure~\ref{globkoef}a), the HCS was flat and was located near
the solar equator $ \sim \pm25^{\circ}$, the GMFS was fragmented
and sharply changed (Figures~\ref{londiag}a, \ref{londiag}b). But
during the minimum of cycle 24, the large-scale two-sector GMFS
existed. Sector structure with large-scale interchanging
positive-polarity and negative-polarity magnetic fields extending
to latitudes of $ \pm40^{\circ}$ on the both sides of the
heliomagnetic equator was observed. CMEs are known to be related
to the heliomagnetic equator and identified with a belt of coronal
helmet streamers (\opencite{Kahler1987};
\opencite{Hundhausen1984}; \opencite{Hundhausen1993};
\opencite{Mendoza1993}). Arcs, connecting opposite-polarity
magnetic fields, separated by the HCS, can be the sources of some
CMEs. Because the HCS was extended to high latitudes, the CMEs
associated with the arcs, were also observed at higher latitudes
than that at the minimum of cycle 23. Such CMEs were faint CMEs.
They had, on average, low speed, width and mass, but rather high
acceleration (Figure~\ref{cme}). Furthermore, the
filaments/prominences are known to locate above the magnetic
polarity inversion line, hence the eruptions of
filaments/prominences will also occur more frequently at higher
latitudes.

Figure~\ref{cme}a  shows that the occurrence rate of CMEs were
higher during the minimum of cycle 24 than that during the minimum
of cycle 23. The daily rate of CMEs was $\sim 3-6$ events at the
minimum of cycle 24 and only $\sim 2-3$ events at the minimum of
cycle 23. A sharp increase in the number of CMEs and their
parameters was observed at the beginning of cycles 23 and 24. The
daily CME rate rose faster at the beginning of cycle 24 than that
of cycle 23 (Figures~\ref{cme}a, \ref{arlatlon}e). For cycle 23,
using CACTus CME catalog, \inlinecite{Robbrecht2009a} found that
the daily CME rate, averaged per year, increased roughly with a
factor of 4 from the solar minimum to maximum, approximately from
2 events during the minimum, to 8 events during the maximum. In
Petrie \citeyear{Petrie2013} based on three independent solar
eruption automated catalogs (not CDAW) CACTus (Computer Aided CME
Tracking project), SEEDS (Solar Eruptive Event Detection System)
and Nobeyama Radioheliograph prominence eruption data, it has been
shown that CME rates and prominence eruptions are both higher for
years 2003-2012 than for 1997-2002. It was concluded that the
result is connected with the weakness of the late cycle 23 polar
field and such an increase was explained by the influence of the
polar field weakening in the late cycle 23 and the beginning of
cycle 24 on the global coronal field structure.
\citeauthor{Vourlidas2010} \citeyear{Vourlidas2010};
\citeyear{Vourlidas2011} showed that the CME mass and mass density
in 2009 were close to their 1996 values but the kinetic energy was
a factor of 1.8 lower and CME velocities were 32\% less than in
1996. \opencite{Robbrecht2009a} compared two catalogs CACTus and
CDAW. In Figure 3 of that article it is shown that the percentage
of narrow CMEs (with the width smaller than $20^{\circ}$),
compared to the total number of CMEs, increases in both the CACTus
(red) and CDAW (blue) data. Thus, these investigations, using
independent data catalogues and methods, provide additional
evidence for the increase in CME numbers and especially in small,
faint events. CME velocities (Figure~\ref{cme}c) increased for
individual CMEs, however, when averaged over CR the growth was not
so noticeable. It means that low-velocity CMEs were dominated at
the beginning of cycle 24. From Figure~\ref{cme}d it is seen that
CME widths have also increased but at the beginning of cycle 24 it
increased much slower than at the beginning of cycle 23, and the
averaged values were also lower. It means that the majority of
CMEs were narrow-width CMEs at the beginning of cycle 24. The
average acceleration of CMEs (Figure~\ref{cme}e) changed little
from the minimum to the maximum in cycle 23, and it even
diminished in cycle 24.

The increase in CME number may have several explanations.
SOHO/LASCO  have observed more narrow CMEs since 2003 by improving
the sensitivity of the instruments. Except that, after 2006, faint
CMEs become easier to identify as overall activity decreases
\cite{Wang2014}. An increase in LASCO telemetry in 2010 resulted
in increase of C2 recorded images from 60 to 104 per day, and
consequently in increase in the detected CMEs \cite{Wang2014}. But
in their detailed study based on CME mass estimates it was shown
that cycle 24 is not only producing fewer CMEs than cycle 23, but
that these CMEs tend to be slower and less massive than those of
cycle 23.

It should be noted that in Petrie \citeyear{Petrie2013} not only
CME data from the LASCO catalogs were used, but also the Nobeyama
Radioheliograph prominence eruption data were analyzed, which are
not associated with the changes of LASCO modes and methods. The
study showed that the number of faint CMEs increased.

Using  ARTEMIS-II, CDAW, SEEDS, and CACTus catalogs,
\citeauthor{Lamy2014} (\citeyear{Lamy2014}) found that all four
catalogs agree on the fact that the CME rate has been increasing
faster than the activity index (SSN and F10.7) during the rising
phase of solar cycle 24. They found also that the difference
between the cycles 23 and 24 minima is conspicuous and
characterized by a broader and fainter equatorial belt in cycle 24
 \cite{Lamy2014}.

Hudson et al. (2014) studding the annual averages of active region
flare productivity, on the  base of NOAA "events" database, have
found that ARs in 2004-2005 (CRs 2012-2038)  had flare
productivity about twice as large as those at other times. This
increase coincide with the peaks in CME number and GMFS changes
(lines 8, 9). It means that some of these CMEs are the result of
that flares.  They noted also that the RHESSI flare counts show an
increase in flare productivity at the C-class level, beginning
from October 2003 \cite{Hudson2014}. According to
\cite{Hudson2010}, the flare/CME ratio diminished by almost an
order of magnitude in the cycle 24 minimum. They suggest that the
cycle 24 minimum corona was relatively easy to disrupt. It was
also found that the global radiance of the K corona was 24\%
fainter during the minimum of solar cycle 24 than during the
minimum of solar cycle 23 \cite{Lamy2014}.

Thus, the increase in the number of weak CMEs is determined not
only by changing of the monitoring regime and CME detection
method, but also by a real increase in the weak events that are
the result of the reducing of the solar global magnetic field
strength.

According to our results, CMEs do not effect the GMFS to a great
degree. Some changes were observed in the border of the structure
only. There were a lot of CMEs during the maximum and the
beginning of the decline phase of cycle 23, but the total GMFS
remained quasi-stable during the rather long time of $\sim1-5$
years. There were 2 large GMFS reorganizations only (in CRs 1970
and 1980) leading to the global redistribution of the
positive-polarity and negative-polarity magnetic fields. Moreover,
during the reorganizations the number of weak, low-energy CMEs
increased (Figure~\ref{cmeparam}). Therefore, CMEs are the
consequence of the GMFS reorganization and not the cause.  In Lio
(Figure 2 in \inlinecite{Liu2009}) it is observed, as  described
in the article, that the CME event caused the local changes in the
shape of the border of the GMFS. But we can see that the general
distribution of positive-polarity and negative-polarity magnetic
fields remained unchanged. Some small short-lived changes in the
GMFS may be caused by some powerful CMEs. Such changes in the GMFS
shape observed during the declining phase, visible in
Figure~\ref{londiag}a, \ref{londiag}a, CRs $\sim2034 \div 2082$ d,
as extensions from the main large-scale structure, may be the
results of some CMEs. But these changes are short in time
$\sim1-3$ CRs. CMEs can have influence on the shape of the GMFS
and change the local shape of the structure only. They do not
change the total distribution of magnetic fields.

It is interesting to note, that during the declining phase, when
the oscillations in CME parameters are more pronounced, the
increase in CME number coincide with the decrease in AR
parameters, such as area, extension and the number of spots in
each AR (Figure~\ref{arlatlon}). The timing of the GMFS
reorganizations are marked by thin vertical lines. The strength of
the magnetic field diminished during these times
(Figures\ref{globkoef}a, \ref{globkoef}a). AR parameters increase
between vertical lines, i.e. between the moments of the GMFS
reorganizations. But CME number increases in times marked by
vertical lines (the times of the GMFS reorganizations). It means
that at least CMEs associated with that ARs are not the cause of
the GMFS reorganizations.

The GMFS is not the cause of CMEs itself, but the structure of the
global magnetic field determines the conditions favorable for
CMEs. Nevertheless, all the CMEs are caused by loss of equilibrium
of the pre-existing magnetic structure. When the GMFS changes the
magnetic field strength both of the magnetic field calculated at
source surface and that measured on the Sun as a star diminished
(Figures~\ref{globkoef}a, \ref{globkoef}2b). But the strength of
the external magnetic field is known to play an important
stabilizing effect on CME eruptions \cite{Schmieder2006}. Even
small instability may lead to an eruption and CMEs became more
frequent when the magnetic "frame" is disrupted. But these CMEs
are faint CMEs. There is no time to accumulate large energy
because even a small instability can cause an eruption. They do
not need to be energetic event to disrupt the overlying magnetic
field structure. The strength of the global magnetic field is low.
When GMFS remains quasi-stable during a long time period, large
long-lived ARs and filaments/prominences can be formed and large
amounts of energy can be accumulated, which is need to disrupt the
existing magnetic configuration in the solar corona, because the
magnetic field of the global magnetic field is high. At the solar
cycle maximum, large long-lived two-sector magnetic structures
were observed. The more stable and larger the structure and higher
the magnetic field strength, the larger and more complex the ARs
that can be formed. Large, long-lived ARs of complex magnetic
field produce intense CMEs. During the solar activity maximum the
increase in CME acceleration was accompanied by increases in
velocity, width, mass and energy. The evolution of the global
magnetic field, both the magnetic field strength and GMFS,
controls the general situation in the Sun's atmosphere as well as
directs the conditions for CME occurrence rate and parameters.

The large structures and redistributions of the GMFS during the
cycle 23 maxima and the beginning of the declining phase coincide
with two peaks in ARs and CMEs activity. It seems that the second
CME peak was developed independently from the AR second peak, and
at the same longitudes and with the same distribution of GMFS as
the first CME peak. According to \inlinecite{Robbrecht2009a} and
\inlinecite{Ramesh2010} the CME second peak shows a delay of 6 to
12 months with respect to the sunspot index. Our result shows the
delay for the second peak was equal to 10 CRs.
\inlinecite{Robbrecht2009a} have proposed that the observed time
delay gives an idea of the time needed to build up the necessary
conditions for CME activity. According to \inlinecite{Du2012}, a
double peak suggests that there are two sources or two decay time
scales. The presence of two peaks may indicate the existence of
two waves of activity, each related to one another, but differing
appreciably in their characteristics \cite{Antalova1965}. In
\citeauthor{Benevolenskaya1998}, (\citeyear{Benevolenskaya1998},
\citeyear{Benevolenskaya2003}), the double peak-structure of the
AR solar cycle was explained as a consequence of the impulsive
nature of the solar activity, or a manifestation of the double
magnetic cycle of two dynamo sources separated in space. It was
proposed that a low-frequency component is generated at the bottom
of the convection zone and produces the 22 year magnetic cycle.
The impulses of solar activity with the period of $1.5-2.5$ years
are formed near the top of the convection zone by reappearing
long-lived complexes of activity, and that these impulses can be
explained by the high-frequency component of the toroidal magnetic
field. The hypothesis of time-space organization of sunspot
activity, like impulses, was considered in
\citeauthor{Gnevyshev1963} (\citeyear{Gnevyshev1963},
\citeyear{Gnevyshev1966}, \citeyear{Gnevyshev1967},
\citeyear{Gnevyshev1977}), \inlinecite{Zolotova2012}. Therefore,
the two-peak structure is a result of the global magnetic cycle
evolution.

The observed periodicities in CMEs could be attributed to a
Rossby-type-wave induced variation of the solar global magnetic
field. We suggest that GMFS and the GMFS reorganizations, as
described above, are a consequence of the changes in the source of
excitation of Rossby waves of different periods in the solar
tachocline. It is proposed that the reconfigurations of the GMFS
are associated with the Rossby wave period changes. Each change in
the oscillation period is associated with the GMFS change, which
result in the destruction of the existing coronal magnetic
structure, and consequently, in the increase in the number of CMEs
and in the changes of their parameter. Therefore, the CMEs of the
time of the quasi-stable GMFS and that of the time of the changes
in wave periods and consequently the reorganization of the GMFS,
have, on average, different parameters.

Thus, the changes in the wave periods and  structural
reorganizations in the global magnetic field can result in the
formation of weak CMEs. This may be the explanation of the
formation of CMEs that are not accompanied by any solar activity
phenomena (flares, filament eruptions, arcs, etc.), for example,
such as, the CME event observed on 2008 June 2 and described in
\inlinecite{Robbrecht2009b}. The event originated along a neutral
line over the quiet Sun. There were no any ARs. The CME was wide
and had low speed ($< 300 \, km \, s^{-1}$). The photospheric
fields were weak ($< 3 G$). The event was classified as
streamer-blowout CME. The probable source region of the CME was
from $30^{\circ}$ to $100^{\circ}$ in CR longitude. According to
our investigation, these features are the characteristic of a CME
associated with GMFS reorganization. From Figure~\ref{londiag} we
can see that there was an abrupt change in the shape of the large
two-sector structure in the longitudinal range
$0^{\circ}\div100^{\circ}$ at that time (CR 2071). The changes
could lead to instability in the solar corona and CME.
\inlinecite{Ma2010} have found that the velocities of the CMEs
without distinct low corona signatures generally range between
$100 \, km \, s^{-1}$ and $300 \, km \, s^{-1}$. They noticed that
some faint changes of the coronal structures could be observed
over the solar limb during such a CME.

\section{Conclusion}   \label{seccon}

The detailed comparison of CME and GMFS cycle evolution shows that
CME activity is not chaotic but it is regulated by evolutionary
changes in the solar global magnetic field. The evolution of the
global magnetic field, both the magnetic field strength and GMFS,
control the general situation in the Sun's atmosphere and direct
the conditions for CME occurrence rate and parameters. The
likelihood of a CME increases rapidly at the moments of GMFS
reorganizations. CMEs do not greatly effect the large-scale
long-lived GMFS. Only some changes in the border of the structure
can be caused by a CME. There is a good relationship between CR
averaged CME number, position angle,   speeds, width, mass, and
energy and global magnetic field strength. Spearman correlation
coefficients are 0.48, 0.40, 0.77, 0.47, 0.42, 0.56 respectively
(the significance level is equal to 0.155).

CME activity has an impulse-like character. During the rising
phases of cycles 23 and 24 the impulses in CMEs coincided with the
impulses in AR and their area. They also coincided with the
reorganizations of the GMFS in a wide range of longitudes, and
with the growth in the new flux emerging. New magnetic fields are
formed, in general, by the emergence of a new magnetic flux
forming new ARs. The coincidence of CME and AR impulses suggests
that the increase in CME activity was associated with ARs. During
the declining phase a new magnetic flux emerged in already
existing ARs. The oscillations in CME parameters, observed during
the declining phase, were the result of the alternation of two
processes. The first one was the emergence of a new magnetic flux
coinciding with the AR area and complicity increase. The CMEs
associated with that process had, on average, higher velocity,
lower width and higher acceleration. The second process was the
structural change in the global magnetic field. The CMEs
associated with the GMFS reorganization were characterized by low
velocity, low acceleration, but higher width. The rate of CMEs
exceeded that of ARs indicating that CMEs were associated with
some other phenomena such as streamers, arcs or
filament/prominence eruptions not associated with ARs. Because the
HCS was extended to the high latitudes during the declining phase
of the cycle 23, the CMEs were also located at higher latitudes.

Our result shows the delay for the second CME peak relative to the
second AR peak was equal to 10 CRs. During each peak, the growth
in CME and AR rate was in the form of individual impulses. The
rate of CMEs closely followed that of ARs only during the rising
phase and the first peak, however, they were very different during
the second peak. The second CME peak was higher than the first
one. Moreover, the new GMFS, associated with the second peak,
appeared in the same longitudes as those of the structure existing
at the time of the first peak, and the periods of oscillations
were the same. The GMFS reorganization during the second CME peak
covered a larger longitudinal interval compared to the first one.

It is suggested that Rossby waves generated in the solar
tachocline result in the observed GMFS. The changes in the periods
of the magnetic Rossby waves result in the reorganizations of the
GMFS which lead to the destruction of the existing coronal
magnetic field structure and consequently the increase of faint
CMEs at the right range of latitudes. The periods of Rossby waves
that can be associated with the GMFS lie in the range $\sim50 \div
1000$ d. The wave periods became shorter from $\sim400 \div 50$ d
from the minimum to the maximum of cycle 23, and they grew to the
minimum of cycle 24 again. The process was not a smooth one, but
it had a form of individual steps. Each period remained
quasi-constant until a new one appeared. The observed GMFS seems
to be a consequence of the excitation of Rossby waves of different
periods. Each Rossby wave period favors a particular GMFS. The
changes in wave periods coincide with the GMFS reorganization and
in the CME location, occurrence rate and parameter changes. The
CME rate and parameters depend on the sharpness of the GMFS
changes, the strength of the global magnetic field and the phase
of a cycle.

These results are important for understanding the global magnetic
field evolution over a solar cycle as well as the complete picture
of CME occurrence rate and parameter changes. Further
investigative research is required to uncover the physical
mechanisms behind the wave generation, and consequently the GMFS
formation and reconfiguration and its influence on CMEs. This
research is of great importance, especially in the context of
solar-terrestrial interaction.

%
 \begin{acks}

This CME catalog is generated and maintained at the CDAW Data
Center by NASA and The Catholic University of America in
cooperation with the Naval Research Laboratory. SOHO is a project
of international cooperation between ESA and NASA.

Wilcox Solar Observatory data used in this study was obtained via
the web site $http://wso.stanford.edu \, at \, 2013:10:19 \,
06:34:02$ PDT courtesy of J.T. Hoeksema. The Wilcox Solar
Observatory is currently supported by NASA.

Information from the Space Weather Prediction Center, Boulder, CO,
National Oceanic and Atmospheric Administration (NOAA), US Dept.
of Commerce were used.

 \end{acks}






\bibliographystyle{spr-mp-sola-cnd} 

\bibliography{bilenko}

\IfFileExists{\jobname.bbl}{} {\typeout{}
\typeout{}}




\end{article}
\end{document}